\documentclass[aps,amssymb,amsmath,pra,twocolumn,showpacs,superscriptaddress]{revtex4-1}

\usepackage{graphicx}
\usepackage{dcolumn}
\usepackage{bm}
\usepackage{subfigure}
\usepackage{color}

\usepackage{amssymb,amsmath,amsfonts,latexsym,graphicx,verbatim}

\usepackage[margin=0.5in]{geometry}

\usepackage[english]{babel}
\usepackage{times}
\usepackage{latexsym}
\usepackage{fancyhdr}
\usepackage{float}
\usepackage{afterpage}
\usepackage{enumitem}
\usepackage{eso-pic, graphicx}

\usepackage{listings}
\usepackage{multirow}
\usepackage[table]{xcolor}

\usepackage{bbm}
\usepackage{upgreek}
\usepackage{amsmath}

\definecolor{Ablue}{rgb}{0.96,0.24,0.00}

\definecolor{Abluetitle}{rgb}{0.,0.24,0.51}
\newcommand{\bluetitle}{\color{Abluetitle}}

\definecolor{orange}{rgb}{0.96,0.24,0.00}

\definecolor{darkred}{rgb}{0.55, 0.0, 0.0}

\definecolor{Gray}{gray}{0.85}
\definecolor{LightCyan}{rgb}{0.88,1,1}
\definecolor{darksalmon}{rgb}{0.91, 0.59, 0.48}
\definecolor{maroon}{cmyk}{0,0.87,0.68,0.32}

\definecolor{mustard}{rgb}{1.0, 0.86, 0.35}
\newcolumntype{a}{>{\columncolor{Gray}}c}
\newcolumntype{b}{>{\columncolor{white}}c}

\usepackage{array}
\newcolumntype{L}[1]{>{\raggedright\let\newline\\\arraybackslash\hspace{0pt}}m{#1}}
\newcolumntype{C}[1]{>{\centering\let\newline\\\arraybackslash\hspace{0pt}}m{#1}}
\newcolumntype{R}[1]{>{\raggedleft\let\newline\\\arraybackslash\hspace{0pt}}m{#1}}

\usepackage[colorlinks=true , citecolor=blue,urlcolor=blue]{hyperref}

\newcommand{\dC}{$^{\circ}$C\:}
\newcommand{\NV}{NV$^{\text{-}}$\:}

\newcommand{\xb}{\beta}

\newcommand{\vxe}{\varepsilon}

\newcommand{\xg}{\gamma}
\newcommand{\xt}{\theta}

\newcommand{\xl}{\lambda}

\newcommand{\xo}{\omega}

\newcommand{\app}{\approx}

\newcommand{\Bp}{B_{\R{pol}}}

\newcommand{\Cs}{{}^{13}\R{C}}

\newcommand{\mC}[0]{\mathcal{C}}

\newcommand{\xD}{\Delta}

\newcommand{\fr}[2]{\frac{#1}{#2}}

\newcommand{\sq}[1]{\sqrt{#1}}

\newcommand{\mH}[0]{\mathcal{H}}

\newcommand{\rt}{\rightarrow}

\newcommand{\beq}{\begin{equation}}
\newcommand{\eeq}{\end{equation}}
                  
\newcommand{\benum}{\begin{enumerate}}
\newcommand{\eenum}{\end{enumerate}}
                    
\newcommand{\bit}{\begin{itemize}}
\newcommand{\eit}{\end{itemize}}

\newcommand{\bea}{\begin{eqnarray}}
\newcommand{\eea}{\end{eqnarray}}

\newcommand{\bev}{\begin{verbatim}}
\newcommand{\eev}{\end{verbatim}}

\newcommand{\non}{\nonumber}

\newcommand{\zt}{\times}

\newcommand{\lb}{\left(}
\newcommand{\rb}{\right)}
\newcommand{\lsb}{\left[}
\newcommand{\rsb}{\right]}

\newcommand{\T}[1]{\textbf{#1}}
\newcommand{\I}[1]{\textit{#1}}
\newcommand{\R}[1]{\textrm{#1}}

\newcommand{\zl}[1]{\label{eqn:#1}}

\newcommand{\zfl}[1]{\protect\label{fig:#1}}
\newcommand{\zfr}[1]{Fig. \ref{fig:#1}}

\newcommand{\expec}[1]{\left\langle #1\right\rangle}

\newcommand{\ba}{\left\{ \begin{array}{lr}}
\newcommand{\ea}{\end{array}\right.}

\definecolor{darkred}{rgb}{0.55, 0.0, 0.0}
\newcommand{\BRd}[1]{\textcolor{red}{#1}} 

\newcommand{\blist}[1]{
 \begin{list}{#1}
 \begin{align}
	 arrow
 \end{align}
 $\checkmark\star
  { \setlength{\itemsep}{3pt}
     \setlength{\parsep}{2pt}
     \setlength{\topsep}{3pt}
     \setlength{\partopsep}{0pt}
     \setlength{\leftmargin}{1em}
     \setlength{\labelwidth}{1em}
     \setlength{\labelsep}{0.5em} } }
\newcommand{\elist}{
  \end{list}  }

\DeclareMathSymbol{\vartheta}{\mathalpha}{letters}{"12}
\DeclareMathSymbol{\theta}{\mathalpha}{letters}{"23}
\DeclareMathSymbol{\phi}{\mathalpha}{letters}{"27}
\DeclareMathSymbol{\varphi}{\mathalpha}{letters}{"1E}

\newcommand{\bef}
{
\begin{figure}[htbp]
\centering
}

\newcommand{\eef}{\end{figure}}

\newcommand{\beginsupplement}{%
        \setcounter{table}{0}
        \renewcommand{\thetable}{S\arabic{table}}%
        \setcounter{figure}{0}
        \renewcommand{\thefigure}{S\arabic{figure}}%
     }

\newcommand{\affA}{ Department of Chemistry, and Materials Science Division Lawrence Berkeley National Laboratory University of California, Berkeley, California 94720, USA.}
\newcommand{\affB}{Department of Physics and CUNY-Graduate Center, CUNY-City College of New York, New York, NY 10031, USA.}

\newcommand{\affD}{Department of Chemical and Biomolecular Engineering, and Materials Science Division Lawrence Berkeley National Laboratory University of California, Berkeley, California 94720, USA.}
\newcommand{\affE}{College of Staten Island, CUNY, 2800 Victory Blvd., Staten Island, New York 10312.}

\newcommand{\affH}{Adámas Nanotechnologies, Inc., 8100 Brownleigh Dr, Suite 120, Raleigh, NC, 27617 USA.}
\newcommand{\affG}{Department of Physics, Ben-Gurion University of the Negev, Be’er-Sheva 8410501, Israel}
\newcommand{\affI}{Gemological Institute of America, 50 W 47th, New York, NY 10036 USA.}
\newcommand{\affJ}{Department of Chemistry, Carnegie Mellon University, 4400 Fifth Avenue, Pittsburgh, PA 15213, USA.}

\begin{document}

\title{\bluetitle{High temperature annealing enhanced diamond $\Cs$ hyperpolarization at room temperature}}
   \author{M. Gierth}\affiliation{\affA}
	 \author{V. Krespach}\affiliation{\affA}
	\author{A. I. Shames}\affiliation{\affG}
	\author{P. Raghavan}\affiliation{\affD}
 \author{E. Druga}\affiliation{\affA}
 \author{N. Nunn}\affiliation{\affH}
\author{M. Torelli}\affiliation{\affH}
\author{R. Nirodi}\affiliation{\affA}
\author{S. Le}\affiliation{\affA}
\author{R. Zhao}\affiliation{\affA}
\author{A. Aguilar}\affiliation{\affA} 
\author{X. Lv}\affiliation{\affA}
\author{M. Shen}\affiliation{\affA}
	\author{C. A. Meriles}\affiliation{\affB}
   \author{J. A. Reimer}\affiliation{\affD}
	 \author{A. Zaitsev}\affiliation{\affE}\affiliation{\affI}
	\author{A. Pines}\affiliation{\affA}
	\author{O. Shenderova}\affiliation{\affH}
	\author{A. Ajoy}\email{ashokaj@berkeley.edu}\affiliation{\affA}\affiliation{\affJ}

\begin{abstract}
Methods of optical dynamic nuclear polarization (DNP) open the door to the replenishable {hyperpolarization} of nuclear spins, boosting their NMR/MRI signature by orders of magnitude. Nanodiamond powder rich in negatively charged Nitrogen Vacancy (\NV) defect centers has recently emerged as one such promising platform, wherein $\Cs$ nuclei can be hyperpolarized through the optically pumped defects completely at room temperature and at low magnetic fields. Given the compelling possibility of relaying this $\Cs$ polarization to nuclei in external liquids, there is an urgent need for the engineered production of highly ``hyperpolarizable'' diamond particles. In this paper, we report on a systematic study of various material dimensions affecting optical $\Cs$\ hyperpolarization in diamond particles -- especially electron irradiation and annealing conditions that drive \NV center formation. We discover surprisingly that diamond annealing at elevated temperatures close to 1720\dC have remarkable effects on the hyperpolarization levels, enhancing them by upto 36-fold over materials annealed through conventional means. We unravel the intriguing material origins of these gains, and demonstrate they arise from a simultaneous improvement in \NV electron relaxation time and coherence time, as well as the reduction of paramagnetic content, and an increase in $\Cs$ relaxation lifetimes. Overall this points to significant recovery of the diamond lattice from radiation damage as a result of the high-temperature annealing. Our work suggests methods for the guided materials production of fluorescent, $\Cs$ hyperpolarized, nanodiamonds and pathways for their use as multi-modal (optical and MRI) imaging and hyperpolarization agents.
\end{abstract}

\maketitle

\T{\I{Introduction:}} -- The development of quantum sensors over the past decade has heralded new opportunities for harnessing the power of quantum technologies in the real world. This includes, for instance, the ultrasensitive probing of magnetic~\cite{Taylor08} and electric fields~\cite{Dolde11}, as well as temperature~\cite{Neumann13,Kucsko13} and inertial rotations~\cite{Ledbetter12,Ajoy12g}. The workhorse platform of this class has been the negatively charged nitrogen-vacancy \NV defect center in diamond, driven by the exquisite room-temperature coherence time of its electronic spin, along with the dual ability to spin polarize \I{and} interrogate it through optical means~\cite{Jelezko06}.

Particulate forms of NV-center endowed diamond ``quantum materials'' portend additional applications that stem from their inherently high surface area and targeting ability through surface functionalization~\cite{Mochalin12,Chipaux18}. Nanodiamond  (ND) particles, for instance, can be serve as \I{deployable} and ``in cell'' quantum sensors~\cite{McGuinness11,Kucsko13,LeSage13,Wu16}. An exciting emerging application involves harnessing lattice $\Cs$ {nuclear} spins in these diamond particles as an auxiliary resource. Particularly attractive is the \I{hyperpolarization} of these $\Cs$ nuclei via the optically pumped NV centers, placing them in {athermal} spin configurations that have greatly enhanced NMR signatures~\cite{Fischer13}. The $\Cs$ nuclear magnetic resonance (NMR) signal can often be enhanced by several orders of magnitude, in a manner completely independent of particle orientation, with the hyperpolarization being carried out at room temperature and with relatively benign resources. Indeed, such optical methods for dynamic nuclear polarization (DNP)~\cite{Carver53,Abragam78} present several advantages over conventional methods involving cryogenic temperatures and large magnetic fields~\cite{Ardenkjaer15}, since the hyperpolarization can be generated \I{replenishably} and under ambient conditions and with modest resources (low optical and microwave powers)~\cite{Ajoy17,Ajoy18pol}. As a representative example, \zfr{laser}A demonstrates small 18$\mu$m microparticles having large nuclear polarization levels $\app$0.23\% in just under a minute of optical pumping. This corresponds to a  gain of $\sim$230 over their Boltzmann polarization at high field (7T) (4$\zt 10^5$ over a polarizing field of 38mT), and to a {million}-fold time acceleration of imaging these particles in MR imaging~\cite{Lv19}.

\begin{figure}[t]
  \centering
  {\includegraphics[width=0.49\textwidth]{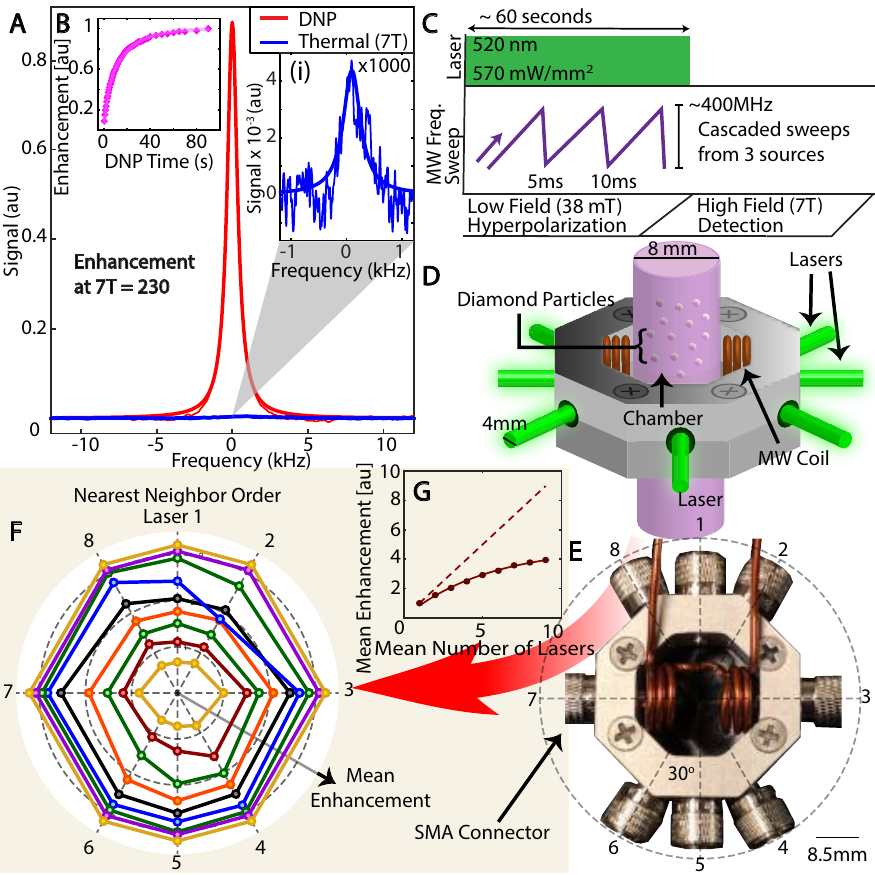}}
  \caption{\textbf{Room temperature optical hyperpolarization.} (A) \I{Typical $\Cs$ hyperpolarization}, shown here for HTA optimized 18$\mu$m diamond particles. Red (blue) line shows the DNP (7T thermal signal, zoomed in inset \I{(i)}), a signal enhancement here $\varepsilon$=230 with respect to 7T. (B) \I{Polarization buildup} curve, showing DNP saturation in $\sim$60s. (C) \I{Schematic of DNP protocol} involving optical pumping at low magnetic fields ($\app$38mT) under simultaneous chirped MW irradiation. $\Cs$ NMR is subsequently detected at 7T.  (D-E) \I{Hyperpolarization setup}.  Laser excitation is via multimode 1mm core optical fibers ($\sim$800mW) arranged in an octagon, with an additional collimated laser applied from the bottom. Beam diameters are $\app$4mm at point of contact with the diamond particle sample, which is carried in a test tube and under water. (F) \I{Hyperpolarization homogeneity} in a 20mg sample upon application of nearest-neighbor (NN) combinations of excitation lasers arranged on the octagonal ring. Polar plot $(r,\xt)$ axes refer to enhancement $\mC_{k,\cdots,k+N}$ and excitation angle of the $k^{\R{th}}$ laser respectively. Approximately circular buildup indicates spatially homogeneous polarization in the sample (\I{lines}). Slight deviations from circularity arise due to experimental imperfections (MW inhomogeneity). (G) \I{DNP scaling} with number of lasers $\expec{\mC_N}$, considering application of an increasing order of NN laser combinations. Sub-linear growth indicates uniform polarization that is not laser power limited.}
\zfl{laser}	
\end{figure}

Several compelling applications open as a result. The optical pumping leads to an associated, simultaneous, bright fluorescence from the \NV spins that is non-blinking and non-bleaching. Harnessing the dual-mode bright nature of these particles in both the optical as well as MR domains opens new avenues for them to serve as imaging agents in disease diagnosis and monitoring~\cite{Aptekar2009,Lv19}. Moreover, nanodiamond particles inherently possess large surface areas ($\sim$57m$^2$/g for 100nm particles~\cite{Costa14}), and along with the long lifetimes of the $\Cs$ spins, allow hyperpolarized surface $\Cs$ nuclei to serve as spin polarization relay channels to external nuclear spins. One could envision, for instance, a liquid coating on the particles being spin-polarized via the surface nuclei through Overhauser contact~\cite{Abrams14}.

While the gains in hyperpolarization can be impressive, experience indicates that material factors play a key role in ultimately setting the obtained $\Cs$ polarization levels. This comes as no surprise; indeed, there has been tremendous progress in the last decade in fabricating diamond particles that are materials-optimized for fluorescence brightness~\cite{Shenderova19,Alkahtani18,Chang08}. This includes diamond growth, e-irradiation, and annealing conditions, that have enabled diamond particles to become some of the brightest and most photostable targetable fluorescent agents available~\cite{Reineck17}. However, analogous optimization of material conditions for optical hyperpolarization is several-fold more challenging.  Apriori, it is uncertain if conditions that produce optically brightest particles (typically few ppm of \NV center concentration and annealing at 850\dC) are also those that make them maximally NMR bright. A rich confluence of factors dictate at once the buildup of $\Cs$ hyperpolarization, and optimized materials are contingent on having high initial polarization and long coherence times on the \NV electron (\I{source}), along with a long lifetime on the $\Cs$ nuclei (\I{target}). These, in turn, are strongly determined by interactions with paramagnetic impurities and lattice distortions. No doubt, these factors are quite different from conditions for radical-induced hyperpolarization at cryogenic conditions~\cite{Casabianca2011}.

In this paper we unravel the various material dimensions that govern optical $\Cs$ hyperpolarization in diamond particles. We accomplish this through combined electron paramagnetic resonance (EPR) and NMR relaxometry studies, along with probes of hyperpolarization buildup and decay behavior for over 30 samples, produced with controllably engineered electron irradiation and annealing conditions. We focus on high-pressure high-temperature (HPHT) diamond particles~\cite{Boudou09}, given their relative ease of production at scale.  To distill effects from each material dimension separately, we consider particles fabricated from identical starting parent materials. While confirming the critical effect of the lattice quality to ultimate $\Cs$ hyperpolarization levels, we discover a surprising new effect upon sample high temperature annealing (HTA) for 15 min at $\sim$1720\dC, which enables a large boost in the hyperpolarization levels, in some cases by over an order of magnitude compared to conventional diamond fabrication. Our data reveals the intriguing origins of these gains, including the crucial role played by paramagnetic impurities in bounding both the electronic as well as the nuclear lifetimes.

\T{\I{High mass $\Cs$ hyperpolarization:}} -- We consider type Ib HPHT particle samples (Hyperion) with identical $\sim$100ppm nitrogen content and compare their relative performance for room temperature DNP at $\Bp\app$38mT, close to the optimal field for hyperpolarization (see \I{Methods}~\cite{SOM}). The particles (all at natural abundance $\Cs$)  are illuminated with 520nm laser light to initialize the \NV electrons to the $m_s=0$ state, and subjected to chirped microwave (MW) irradiation over the \NV EPR spectrum to transfer polarization to the $\Cs$ lattice (see \zfr{laser})~\cite{Ajoy17, Ajoy18}. Note that, in this paper, we will refer to the NV center as being in its negatively charged state, and ignore effects of charge dynamics. \zfr{laser}B shows typical hyperpolarization buildup curve, saturating in $\sim$90s of optical pumping. Polarization transfer under the MW sweeps (\zfr{laser}C) proceeds through a sequence of rotating frame Landau-Zener transitions~\cite{Zangara18} to relatively weakly coupled $\Cs$ nuclei, and subsequent spin-diffusion serves to homogenize the polarization in the bulk lattice. Interactions of the $\Cs$ nuclei with other lattice spins, especially paramagnetic impurities, contribute to the leakage of polarization to unmeasurable degrees of freedom that effectively manifests as $T_1$ relaxation, bounding the overall polarization level. 

Experimentally, the hyperpolarized $\Cs$ NMR signals are measured at high field (7T), accomplished by rapidly shuttling the sample prior to measurement.  Benchmarking the hyperpolarization level of each sample against its counterpart 7T Boltzmann value is unwieldy and time-consuming, since a typical thermal measurement takes over 30h for discernible SNR, limited by low sample filling factor ($\lesssim10^{-4}$) in detection. Instead here, we adopt an alternate strategy. For a fair comparison, we mass normalize the hyperpolarization signals from the various samples and benchmark each sample against the unit mass thermal NMR signal at 7T.  We report these DNP {enhancement} factors, henceforth labeled $\varepsilon$, which correspond to absolute polarization levels $\app 0.1\varepsilon\%$. 

Technically, to ensure the entire mass (5-30mg) of the diamond particles (occupying $\lesssim$18mm$^3$ volume immersed in water) is polarized uniformly, we employ a laser illumination geometry that serves to penetrate all particle facets and ensure homogenous powder polarization in spite of scattering and penetration losses. This is accomplished by using eight 800mW fiber-coupled lasers arranged along an octagonal ring, along with another collimated laser that excites the sample from the bottom, approximating a hemispherical excitation pattern (see \zfr{laser}D-E). The optical excitation ($\sim$570mW/mm$^2$) is minimally obstructed by the MW coil, which is a split-solenoid of 9mm diameter (see \zfr{laser}E), and with an exceedingly low MW power density ($\sim$2.5mW/mm$^3$). To quantify the effectiveness of this excitation geometry consider that, for instance, $\mC_{k_1,k_2}$ refers to the obtained DNP enhancement employing laser numbered $k_1$ and $k_2$ on the octagonal ring together. We now consider (\zfr{laser}F) the DNP signal under \I{nearest-neighbor} (NN) combinations of the laser beams, $\mC_{k}, \mC_{k,k+1},\cdots,\mC_{k,\cdots,k+N}$, where $k\in [1,8]$ is the laser number on the ring (see \zfr{laser}E), and $N$ refers to the NN order. The inner yellow line in \zfr{laser}F for instance refers to $N=1$ (single lasers), the red line to $N=2$ (NN pairs), and the green points to $N=3$ (NN triples).  DNP levels here are shown in a polar plot, where the radius indicates the obtained signals, and the angular coordinate represents laser $k$ above. The approximately circular ring-like patterns in \zfr{laser}F demonstrates the spatially homogeneous buildup of polarization in the powder samples. The slight deviations from circularity can be ascribed to MW inhomogeneity due to the MW coil being off-center with respect to the lasers.  \zfr{laser}G finally illustrates the mean signal obtained for each NN laser order, $\expec{\mC_N}=\expec{\mC_{k,\cdots,k+N}} = \fr{1}{8}\sum_j \mC_{k_j,\cdots,k_j+N}$; the sub-linear scaling in \zfr{laser}G demonstrates that the DNP in the $\app$20mg mass is not laser-limited. In the Supplementary Information~\cite{SOM}, we also study the correlation between pairs of lasers situated on the octagonal ring, measuring $\expec{C_{k,k+n}} = \fr{1}{8}\sum_j \mC_{k_j,k_{j+n}}/(\mC_{k_j} + \mC_{k_{j+n}})$, the mean value of the DNP enhancement on applying two lasers separated by $n$ simultaneously, and normalized by the effect of them both applied separately. We find that (see ~\cite{SOM}), to a good approximation, the different lasers contribute to hyperpolarization in different parts of the sample volume, and the near hemispherical excitation leads to uniform polarization buildup.

\begin{figure}[t]
  \centering
  {\includegraphics[width=0.49\textwidth]{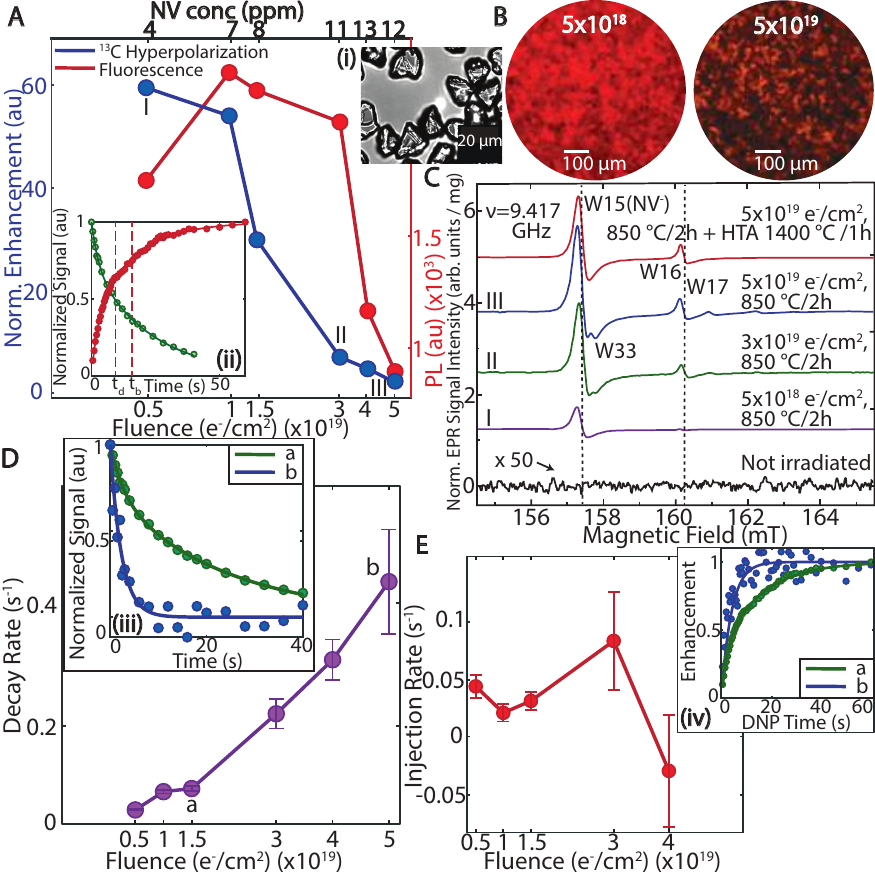}}
  \caption{\T{DNP scaling with electron irradiation dose}. (A) \I{Mass normalized} hyperpolarization enhancements for 18$\mu$m particles at $\sim$38mT (blue line) showing steep decrease at high irradiation fluences. Corresponding \NV concentrations are denoted on upper axis (bold). Red line shows optical fluorescence and displays a similar trend. \I{Inset (i)}: particle micrograph showing particle size 18$\mu$m$\pm$3$\mu$m. \I{Inset (ii)}: representative polarization buildup (red) and decay (green) curves. Buildup $t_b$ and decay $t_d$ times are denoted by respective $1/e$ intercepts. (B) \I{Florescence micrographs} of representative samples marked in (A) demonstrating differences in optical brightness. (C) \I{X-band EPR spectra}. Growth of \NV center (W15) peaks at larger electron fluence doses is associated with an increase in additional triplet paramagnetic defects (here triplet centers W33, W16-17). (see also \zfr{EPR}) (D) \I{Polarization decay rates} for the samples considered in (A). \I{Inset (iii):} representative polarization decay curves for low and high electron dose samples. (E) \I{Polarization injection rate} extracted from the buildup curves. Deleterious effects on both the polarization injection and decay rates at high electron fluences lead to reduced hyperpolarization efficiency. \I{Inset (iv):} polarization buildup curves for representative samples in (D).}
\zfl{NVconc}	
\end{figure}

\T{\I{Effects of increasing electron fluence}} -- First, considering the effects of particle size by studying particles obtained through milling identical starting materials, we find a decrease in hyperpolarization levels at small particle sizes, 100nm particles performing worse by about an order of magnitude compared to 18$\mu$m particles (see Supplementary Information~\cite{SOM}). However since this is strongly conditioned on the harder to control material degradation due to milling, we will consider a more systematic study elsewhere. Instead, for clarity, we focus here almost completely on diamond particles that are identically milled to a uniform size 18$\pm$3$\mu$m, all starting from the same parent material.  we consider in \zfr{NVconc} samples prepared under varying fluences of electron irradiation and \I{standard} annealing conditions (850\dC for 2hrs), both of which can be precisely controlled. Naively, the increasing fluence results in an increasing \NV center concentration that should seed a greater polarization in the $\Cs$ lattice. Practically, however, this is associated with concomitant lattice damage, as well as an increased paramagnetic defect concentration, resulting in a decrease in $\Cs$ $T_1$ times, and manifests as an inherent tradeoff in the hyperpolarization levels with fluence. This is evident in the experiments in \zfr{NVconc}A where we observe a DNP optimum at $\app5\zt 10^{18}$e/cm$^2$, which we estimate from EPR data to correspond to an \NV concentration of $\sim$4ppm (upper axis). Furthermore, the DNP enhancements correlate with the optical brightness of the particles (red line in \zfr{NVconc}A), which also decreases at higher fluences, here due to the generation of paramagnetic optical traps. \zfr{NVconc}B, for instance, shows representative micrographs taken under a 617nm long-pass filter highlighting the relative particle brightness.

Continuous wave X-band (9.4 GHz) EPR measurements (at room temperature) highlight underlying lattice conditions responsible (\zfr{NVconc}C). We calibrate $g$-factors and densities $N_s$ of paramagnetic $S$ = 1/2 species against a reference sample of purified detonation nanodiamond (ND) powder with g = 2.0028(2) and $N_s$ = 6.3$\zt$10$^{19}$spins/g~\cite{Shames02}. Spectra recorded in the half-field ($g$ = 4) region allow reliable quantification of \NV(W15) triplet centers, tracking EPR lines originating from forbidden ($\xD m_s$ = 2) transitions between triplet state Zeeman levels. \zfr{NVconc}C illustrates that increasing electron fluence leads to the growth of g = 4.274(5) \NV center signals; and while at low fluences (5$\zt$10$^8$e/cm$^2$), the NVs dominate the spectrum, higher fluences are associated with the appearance and strengthening of additional signals, here attributable to W16-W18 and W33 {triplet} defect centers. We emphasize that for clarity, the spectra in \zfr{NVconc}C focus on the spectral region close to the \NV peaks, and do not show the primary lattice paramagnetic defects which also change in concentration (see \zfr{EPR}C and \cite{SOM}).

\begin{figure}[t]
  \centering
  {\includegraphics[width=0.42\textwidth]{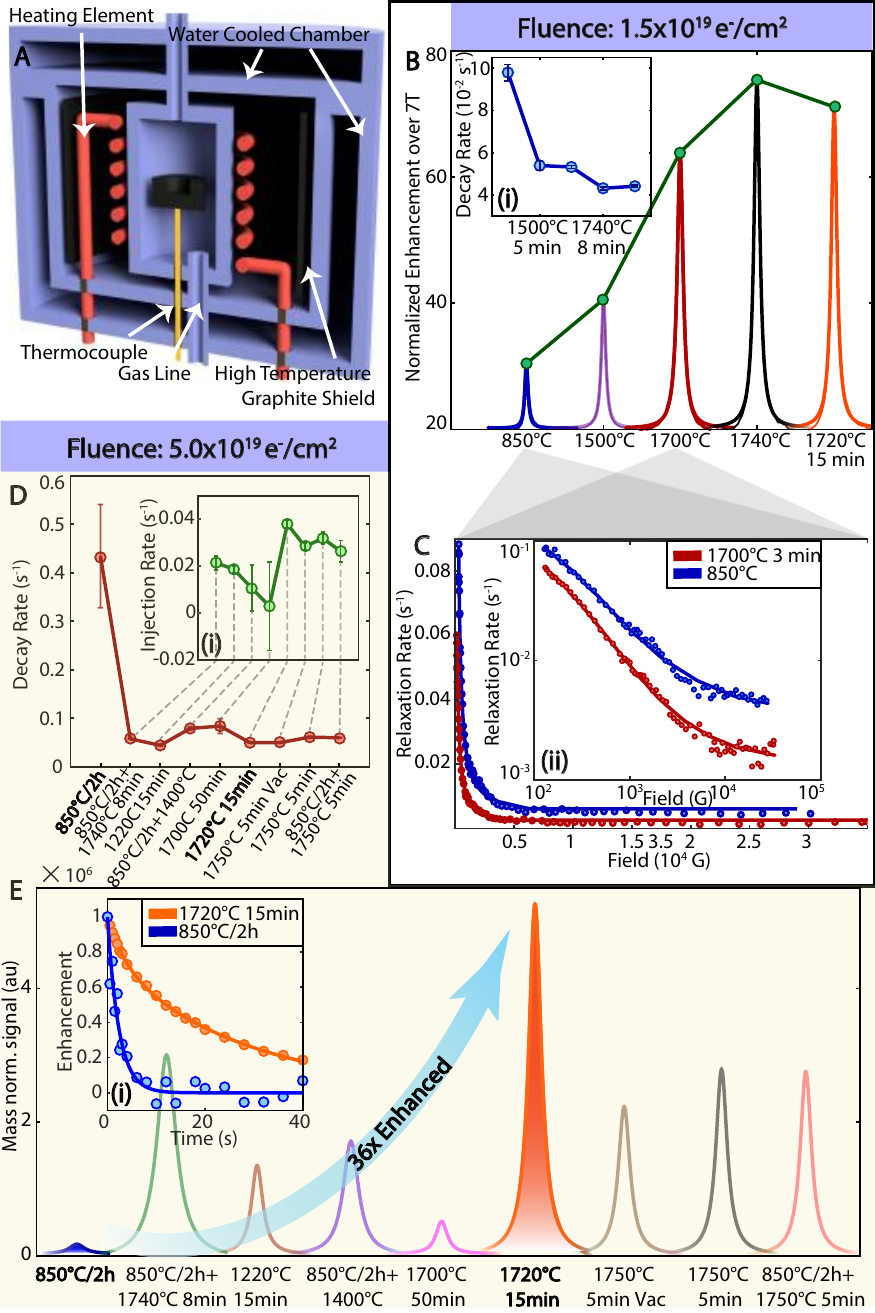}}
  \caption{\T{High thermal annealing (HTA) hyperpolarization gains.} (A) \I{Home-built HTA furnace} consisting of $\sim$1cm$^3$ sample container and ability for precise control of temperature, annealing time and environmental conditions. (B) \I{Mass-normalized $\Cs$ NMR spectra} for samples produced with fluence of $D_1$=1.5$\zt$10$^{19}$/cm$^2$, corresponding close to brightest particles (8ppm \NV concentration) in \zfr{NVconc}A. HTA causes a $\sim$3x increase in hyperpolarization levels compared to conventional annealing conditions (850\dC). \I{Inset (i)}: polarization decay rates, indicating that HTA conditions produce an $\sim$2.5x increase in $\Cs$ relaxation lifetimes. (C) \I{$\Cs$ $T_1$ Relaxometry} reveals the origins of this enhanced lifetime for a representative HTA sample. We map $\Cs$ relaxation rates $R_1(B)$ over a wide field range 100mT-3T. (\I{Inset:} data on log scale). Solid lines are fits to Tsallian functions (see ~\cite{SOM}). We observe a sharp narrowing of relaxometry profiles under HTA, manifesting in a increase in nuclear $T_1$ at all fields. (D-E) \I{Impact of HTA on high fluence samples} with $D_2$=5$\zt$10$^{19}$/cm$^2$, originally corresponding to the worst performing sample in \zfr{NVconc}A. (D) \I{Upper panel:} Measured relaxation and injection rates under different HTA conditions, showing marked improvement in $\Cs$ lifetimes. \I{Inset:} Extracted polarization injection rates, showing improvement under sample HTA. (E) \I{Lower panel:} Mass-normalized spectra reveals a dramatic $\sim$36x increase in the hyperpolarization enhancements under HTA over conventional annealing (blue arrow). While a wide range of HTA conditions can improve performance, optimal HTA conditions are found to be at $\app$ 1720\dC for 15min. \I{Inset:} Comparative $\Cs$ relaxation decays at 38mT, showing wide contrast in relaxation times with and without HTA.}
\zfl{RTA}	
\end{figure}

The presence of paramagnetic spins bottlenecks hyperpolarization buildup in the high e-fluence samples. Indeed, the saturation hyperpolarization values reflect a dynamic equilibrium between \NV-induced polarization injection into the lattice $\Cs$ nuclei, and its inherent decay due to nuclear $T_1$ processes. If $\xb_i$ and $\xb_d$ denote the (assumed monoexponential) polarization injection and decay rates respectively, the buildup curve (\zfr{NVconc}A (ii)) has the functional form, $\vxe(t) = \fr{\xb_i}{\xb_i+\xb_d}\lsb 1 – e^{-(\xb_i+\xb_d)t}\rsb$. Hence for each sample, measurement of the polarization buildup and decay curves allows an independent estimation of injection and decay rates. Both parameters provide valuable insight into the material conditions that affect hyperpolarization levels; if for instance $\xb_d$ is large, polarization saturates at a low value in spite of high \NV concentrations. The buildup and decay curves are however, in general, weakly bi-exponential (see \zfr{NVconc}A) on account of disparate behavior between directly \NV-coupled $\Cs$ and weaker bulk nuclei. We can nonetheless, to a good approximation, quantify the buildup and decay rate constants (and error bars) through the inverse $\lb 1-1/e\rb$ and $1/e$ intercepts of the fitted lines respectively, depicted as $t_b$, $t_d$ in \zfr{NVconc}A. This provides a faithful reflection of the underlying injection rates, although carrying larger error in the limit of $T_{1}\rt 0$.

The extracted $\Cs$ relaxation rates grow steeply for the high e-fluence samples (\zfr{NVconc}D), and as the representative decay curves from low and high fluence samples in \zfr{NVconc}D(iii) indicate, the differential in $T_1$ values can be as large as 5 times. We emphasize that the nuclear $T_1$ (spin-lattice) relaxation here is not phonon mediated; instead, at the operational hyperpolarization fields, it originates predominantly from stochastic spin-flipping noise produced at $\Cs$ sites from lattice paramagnetic electrons~\cite{Ajoy19relax}. This mechanism is dominant since the nuclear Larmor frequency, $\xo_L=\xg_n\Bp$ (for instance $\app$380kHz at 38mT) can lie \I{within} the dipolar-broadened EPR linewidth. Specifically, the line broadening $\expec{d_{ee}}\app \xg_e\sq{\fr{8}{\pi}}\sq{M_{2e}}\:$ [Hz] $\app$10.5$P_e$ [mG], scales approximately linearly with electron concentration $P_e$~\cite{Reynhardt03a}, where $M_{2e}$ is the second moment of the electronic spectra~\cite{Abragam61}
$
M_{2e} = \fr{9}{20} (g\mu_B)^2\fr{1}{\expec{r_e}^6},
$
and $g\app 2$ is the electron $g$-factor, $\mu_B=9.27\zt10^{-21}$ erg/G is the Bohr magneton, the 
inter-spin distance is $\expec{r_e}=\lb 3/4\pi \ln 2\rb^{1/3}N_e^{-1/3}$, and  $N_e=(4\zt 10^{-6}P_e)/a^3$[m$^{-3}$] is the electronic concentration in inverse volume units, and $a$ = 0.35 nm the lattice spacing in diamond~\cite{Reynhardt03a}. Note that here we measure indirectly the cumulative effect of the paramagnetic spins on the $\Cs$ nuclei and are agnostic to their exact lattice origin. The steep increase in relaxation rate in \zfr{NVconc}D therefore reflects the deleterious increase in paramagnetic content on the $\Cs$ nuclei, effectively increasing $\expec{d_{ee}}$ (see also \zfr{RTA}C).

Similarly, turning our attention to the extracted polarization injection rates (\zfr{NVconc}E), we observe a saturation and decrease with increasing electron fluence. This is somewhat counterintutive, since naively an increase in \NV concentration should result in a greater number of sources seeding polarization into the $\Cs$ lattice. However the NV-$\Cs$ polarization transfer is a coherent process, and \zfr{NVconc}E strongly suggests that the transfer efficiency \I{per} MW sweep event is decreased in the higher fluence samples. This is potentially due to the reduction in \NV center coherence time $T_{2e}$ from interactions with the surrounding paramagnetic spin bath~\cite{Bauch18}, as well as charge noise from lattice ionic nitrogen species~\cite{Kim15}. \zfr{NVconc} therefore  illustrates that while would one like to electron irradiate samples to maximize the number of \NV centers, this comes at a steep cost of lattice damage and paramagnetic impurity content, and strongly countervails against $\Cs$ hyperpolarization buildup. 

\begin{figure*}[t]
  \centering
  {\includegraphics[width=1\textwidth]{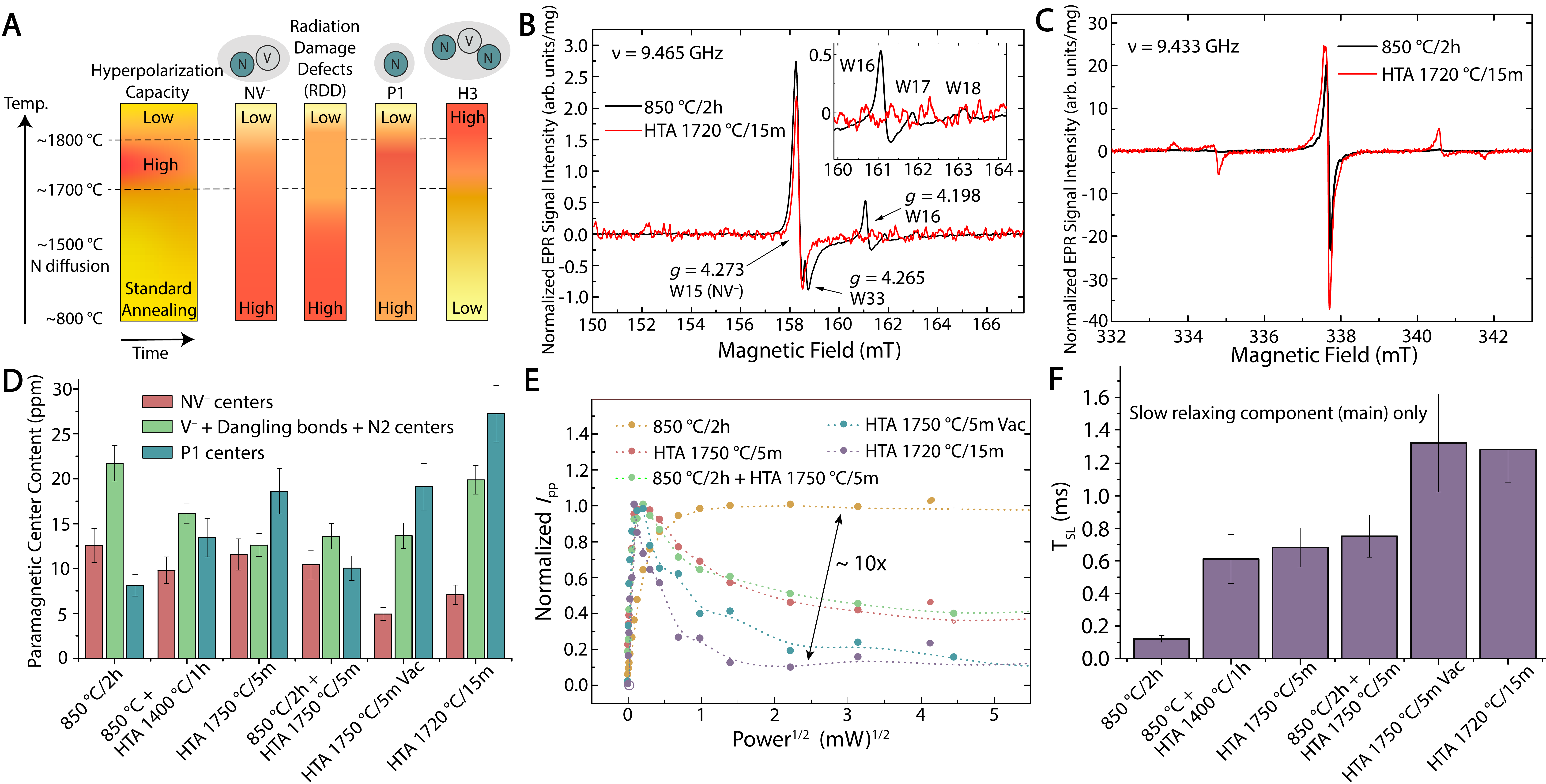}}
\caption{\T{EPR spectra at X-band}, revealing lattice origins of the HTA-driven hyperpolarization gains. (A) \I{Schematic of defect center evolution} typical to HTA with changes in annealing time and temperature (\I{axes}). We target a HTA condition between 1700-1800\dC (dashed lines). (B) \I{EPR spectra} at high electron dose ($D_2$= 5$\zt$10$^{19}$e/cm$^2$), displaying \NV centers and spectrally closeby paramagnetic impurities (W33) centers (see also ~\cite{SOM}). HTA (\I{red}) suppresses the W33 and W16-W18 centers in its conventionally annealed counterpart (\I{black}). This manifests in EPR spectra similar to samples irradiated at low dose (see \zfr{NVconc}C). \I{Inset:} Zoom in on the accompanying triplet defect centers. (C) \I{P1 center EPR spectra} shows an increase in P1 concentration under HTA, indicating lattice reorganization of nitrogen from N$^{+}$ to P1 (total nitrogen content in all samples is identical $\app$100ppm).   (D) \I{Extracted concentrations} of the $S$=1/2,3/2 and $S$=1 paramagnetic defects from the EPR spectra. (E-F) \I{\NV spin-lattice relaxation times} under several HTA conditions. (E) \I{Raw data MW saturation curves} measured on the low field allowed transition in EPR spectra of \NV centers (data for forbidden transitions in ~\cite{SOM}). (F) \I{Extracted $T_{1e}$ times} through exponential fit of the saturation curves in (E) (see ~\cite{SOM} for similar data on forbidden transitions). There is a boost of about an order of magnitude in electron $T_{1e}$ in the 1720\dC HTA sample compared to conventional annealing.}
\zfl{EPR}	
\end{figure*}

\T{\I{High temperature annealing enabled DNP gains}} --  The samples considered so far were subjected to standard annealing, conditions that maximize optical fluorescence luminosity. On the other hand, annealing at higher temperatures could have a conciliatory role by relieving lattice disorder, due to the evolution and reorganization of defects~\cite{Jones15}. Motivated by this, we consider the effect of ultra-high temperature annealing on the hyperpolarization levels, employing a water-cooled annealing furnace (\zfr{RTA}A) that allows precise temperature control as well as rapid temperature rise times (see \I{Methods}). While the annealing process constitutes a multi-dimensional parameter space dependent on temperature, annealing time, and sample environment conditions; here we consider an important subset of these conditions, focusing on annealing in the 1700-1800\dC range, where \NV centers can be preserved while paramagnetic content decreased~\cite{Dei19} (\zfr{EPR}A).

We concentrate here on two representative classes of samples, those irradiated at $\T{\I{(i)}}$ $D_1$=1.5$\zt$10$^{19}$e/cm$^2$ (\zfr{RTA}B-C) that were optimally fluorescent in \zfr{NVconc}C, and $\T{\I{(ii)}}$ at the maximum fluence level $D_2$=5$\zt$10$^{19}$e/cm$^2$ (\zfr{RTA}D-E), possessing close to the largest \NV center concentration. Our observations (see \zfr{RTA}B,E showing mass normalized $\Cs$ spectra) indicate that HTA in the 1700-1800\dC range can consistently lead to large improvements in the hyperpolarization levels over conventional annealing at 850\dC. For the $D_1$ samples (\zfr{RTA}B), HTA boosts polarization levels by $\sim$3x, while for the $D_2$ samples (\zfr{RTA}E), originally close to worst performing in \zfr{NVconc}C, we find a large 36-fold increase (arrow in \zfr{RTA}E) in the magnitude of DNP enhancement levels under optimal 1720\dC HTA (shaded in \zfr{RTA}E). This allows these samples to leapfrog to the best overall for hyperpolarization despite initially showing very poor performance. If employed in $\Cs$ hyperpolarized MR imaging~\cite{Lv19}, this would correspond to a substantial acceleration just by changes in the annealing conditions. These large gains are at once both surprising and technologically significant, since annealing at 1720\dC is not substantially more technically challenging than at 850\dC (\zfr{RTA}A)~\cite{eaton17}, and time and temperature control opens new parameter spaces for lattice defect manipulation.

We notice moreover that a wide range of temperature conditions, anywhere in the 1200-1800\dC range, can yield substantial DNP improvements. Interestingly, short-time \I{post}-annealing a sample originally annealed at 850\dC at 1400\dC for 1hr can already increase DNP levels by $\sim$3x (second trace in \zfr{RTA}E). The data also reveals the importance of relatively \I{rapid} HTA, extended annealing for instance 50min can lead to significantly degraded performance (fifth trace in \zfr{RTA}E). We observe that the exact environment of the annealing does not matter strongly, although a H$_2$ atmosphere is marginally better and can also serve to reduce diamond graphitization~\cite{Dei19} (\zfr{RTA}E). To put in perspective the best HTA performance under the two electron doses, illustrating that $D_2$ samples outperform the lower fluence samples ($\gtrsim$1.5x greater than any of the samples considered here), and indicating that the high \NV concentration now can be usefully harnessed. The best performance corresponds to a bulk $\Cs$ polarization $\sim$0.3\% (\zfr{laser}A) and is the highest reported optical hyperpolarization level on crushed particles $<$20$\mu$m in size. Finally, the data in \zfr{RTA}B-D strongly indicate that these DNP signal boosts can be explained atleast partially by the strong ($\sim$ 3-5x) increase in $\Cs$ relaxation times as a result of HTA. This is most evident in the representative polarization decay curves in the inset of \zfr{RTA}E. We also observe a measurable increase in the polarization injection rates under HTA (inset in \zfr{RTA}D).

We now employ a combination of NMR relaxometry and quantitative EPR data to unravel lattice origins of these HTA-driven gains. Considering first the effect on increasing $\Cs$ relaxation times, we find that this increase occurs across all magnetic fields (\zfr{RTA}C). To illustrate this, we use a recently developed NMR relaxometry method~\cite{Ajoy19relax} to quantify $R_1(B)=1/T_1$, mapping the nuclear relaxation rates with magnetic field. Experiments are performed with a home-built field cycler device~\cite{Ajoyinstrument18}, rapidly shuttling the sample to the field of interest where the spins are allowed to relax for a waiting period $t$, and subsequently detecting the polarization decay at 7T. Stepping over values of $t$ unravels the \I{full} $T_1$ decay curve at every field point, here with high precision (50 field points) over 100mT-3T (\zfr{RTA}C). In practice, we further accelerate the data acquisition using a dimension-reduction protocol outlined in Ref.~\cite{Ajoy19relax}. Fits to an exponential decay $\sim\exp\lb-t/T1\rb$ allow extraction of field-dependent rates $R_1(B)$, which display characteristic {step-like} dependencies with field (shown in a linear scale in \zfr{RTA}C and log scale in the inset), growing steeply by over 2 orders of magnitude at low field and saturating beyond $\sim$100mT to a relatively long $T_1$ value ($\app$10min). The relaxometry profiles widths $\sim\fr{\expec{d_{ee}}}{2\xg_n}$ report on the interactions of the $\Cs$ nuclei with lattice paramagnetic spins~\cite{Ajoy19relax}. In these experiments, we cannot resolve differences between the exact paramagnetic spin species, but instead measure their net effect on the $\Cs$ nuclei. The effects of HTA are now clearly evident (see \zfr{RTA}C); it leads to markedly modified NMR relaxometry profiles, narrowing them (over 20\%) even for the low fluence samples ($D_1$=1.5$\zt$10$^{19}$e/cm$^2$) considered here. As a consequence, at any hyperpolarizing field, there is an increased time over which the $\Cs$ polarization can buildup in the lattice, and consequently an enhanced saturation polarization level.

Complimentary aspects are revealed by X-band EPR spectra (see \zfr{EPR}), where we focus on the samples irradiated with the higher fluence $D_2$= 5$\zt$10$^{19}$e/cm$^2$, since they present the largest relative gain in DNP levels in \zfr{RTA}.  \zfr{EPR}A schematically displays the expected propagation of lattice defects under HTA conditions, and indicates the possibility of reducing radiation induced damage. We speculate that the HTA gains occur beyond the temperature corresponding to the offset of N diffusion $\sim$1500\dC, since some of the triplet defects might be associated with complexes of N and vacancies (\zfr{EPR}A). Consider first \zfr{NVconc}C the case when the starting sample is annealed additionally at 1400\dC. The elevated temperature now causes the complete disappearance of W17-18 and W33 defects (top trace in \zfr{NVconc}C), although at the cost of some reduction of \NV concentration.  The W16 content, on the other hand, remains practically unharmed in this case. Moving on to high temperature annealing in the 1700-1800\dC range (dashed lines in \zfr{EPR}A), the effect on the triplet defects’ content is even more pronounced – (red trace in \zfr{EPR}B); most evident in the inset in \zfr{EPR}B that zooms into the spectral region near g=4.196. The HTA treatment at 1720\dC completely eliminates all other e-beam induced triplet centers while only slightly affecting the \NV content. Indeed, a fraction of the N atoms congregate into groups of 2-3 N atoms and combine with vacancies forming \I{non}-paramagnetic H3 and N3 color centers (\zfr{EPR}A). It is remarkable that the high \NV density samples upon HTA treatment begin to resemble the triplet defect free spectra (second-to-last trace in \zfr{NVconc}C) that characterize the samples at low electron fluence. The high \NV concentration ($\sim$4ppm greater than \zfr{NVconc}C) is now available free of adverse triplet paramagnetic content~\cite{Goss04}. In the Supplementary Information~\cite{SOM} we present similar EPR spectra under a range of annealing conditions, showing progressive lattice quality improvement with annealing temperature. 

Annealing also affects the pool of primary defects, evident in the EPR spectra in \zfr{EPR}C (see~\cite{SOM}): increasing the remanent substitutional nitrogen defects – P1 centers (blue bars in \zfr{EPR}D), and modestly decreasing the content of negatively charged vacancies V$^{\text{-}}$, dangling bonds and N2 centers, deformation-produced dislocations decorated by nitrogen~\cite{Mineeva09} (green bars in \zfr{EPR}D). To explain this, we note that while the diamond samples considered here all have identical $\sim$100ppm of total nitrogen content, however a predominant proportion manifest as N$^+$ ions that are not paramagnetic and hence are invisible in EPR. The increase in the P1 center concentration upon HTA, while a somewhat surprising observation, reflects the lattice reorganization from N$^+$ to P1 at high temperature. This likely serves to reduce the charge noise seen by the \NV centers and can potentially contribute to their increased coherence times as well as $T_{1e}$ lifetimes. 

Indeed the most dramatic change upon HTA is found in the spin-lattice relaxation times of the \NV centers (\zfr{EPR}E-F). We note that \NV $T_{1e}$ is critical because at the relatively low laser illumination levels and slow MW sweeps we employ (200/s), it dictates the level to which the \NV spins are polarized, ultimately bounds the $\Cs$ nuclear polarization~\cite{Ajoy17,Jarmola12}. $T_{1e}$ times here are estimated by fitting progressive microwave saturation curves obtained for low-field allowed and forbidden (g = 4.274) lines of the \NV EPR spectra, to a two-component model and assuming that the slow and fast relaxing entities originate form the same species. The saturation curves in \zfr{EPR}E (and ~\cite{SOM}) indicate that faster relaxing species become saturated at higher MW power levels whereas sharp saturation peaks correspond to slower relaxing spins. It is evident in \zfr{EPR}E that HTA causes prolongation of spin-lattice relaxation times (see ~\cite{SOM} for related data on the forbidden transitions). The longest $T_{1e}$ values, found for HTA 1750\dC (vacuum) and HTA 1720\dC treatments -– and are about an order of magnitude longer than those conventionally 850\dC  annealed (\zfr{EPR}F). 

It emerges therefore from \zfr{RTA} and \zfr{EPR} that the HTA enabled DNP gains arise from an interplay between increasing $\Cs$ relaxation time by $\sim$3-5x, increasing \NV center $T_{1e}$ by $\sim$10x, as well a suppression in adverse triplet paramagnetic impurity content, all of which can be traced to HTA-driven “\I{healing}” of the diamond lattice disorder caused by radiation damage. Indeed, the best sample studied (HTA at 1720\dC) stands out with respect to its conventionally annealed counterpart; and is characterized by both a relatively high (7ppm) \NV content and a long $>$1ms electron $T_{1e}$, as well as a complete absence of other e-beam induced triplet centers except for NVs. In contrast, the original 850\dC annealed sample had the highest content of remanent paramagnetic vacancies and secondary triplet centers, as well as the shortest $T_{1e}$ values and only moderately higher \NV content ($\sim$12ppm). We anticipate that larger gains are possible through further optimization of the HTA process, and while demonstrated here for 18$\mu$m particles, we expect these would gains also be translatable to nanodiamond samples. 

\begin{figure}[t]
  \centering
  {\includegraphics[width=0.45\textwidth]{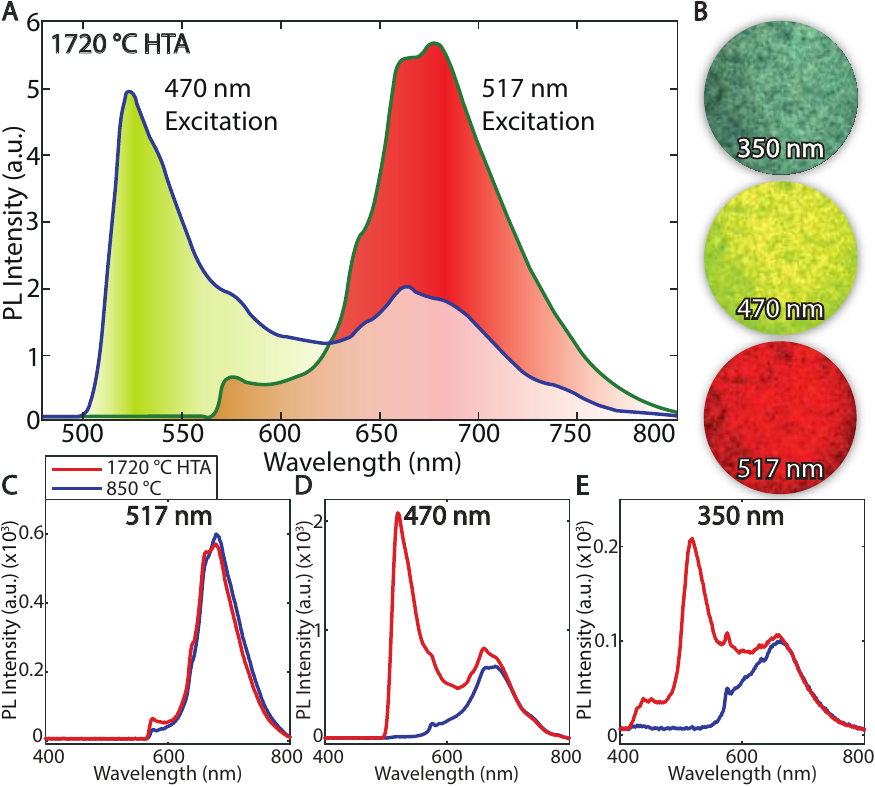}}
\caption{\T{HTA driven multicolor diamond fluorescence.} (A) PL spectra measured for the HTA 1720\dC samples in \zfr{RTA}E studied under 470nm and 517nm excitation wavelengths respectively, showing widely discriminated spectra. Shading indicates predominant florescence color. (B) \I{Multi-color florescence particle micrographs} here for HTA 1720\dC samples. We employ excitations at 350nm (UV), 470nm (blue) and 517nm (green), and image under 405nm, 488nm, 561nm long-pass filters respectively. (C-E) \I{Comparison} in PL spectra between HTA (\I{red}) and conventionally (\I{blue}) annealed samples, showing diversity of fluorescence color with excitation wavelength in the HTA case due to formation of nitrogen related defects. Conventional samples, in contrast, only predominantly fluoresce red.}
\zfl{multicolor}	
\end{figure}

\T{\I{Applications in multi-modal imaging:}} -- Apart from enhanced hyperpolarization, HTA fortuitously also endows the particles with the surprising property of being fluorescent in more than one “color” under different excitation wavelengths~\cite{Dei19}. This effect arises due to the formation of \NV, H3 and N3 centers, and mixtures thereof, conditional on and controllable via the annealing temperature. \zfr{multicolor}B for instance demonstrates this for the best samples for hyperpolarization (HTA at 1720\dC), where the same particles glow in the dark-green, light-green or red based on the UV (350nm), blue (470nm) or green (532nm) excitation wavelengths. \zfr{multicolor}A shows the measured PL spectra,  displaying a wide separation in the dominant fluorescent wavelengths. \zfr{multicolor}C-E highlights the strong differences to the conventionally annealed (850\dC) sample, wherein the PL is always predominantly in the red. We clarify however that HTA does not measurably enhance PL brightness from their values in \zfr{NVconc}B, indicating that while DNP is more contingent strongly on the underlying lattice quality, while PL is more strongly conditioned on fluorescence quenching charge traps~\cite{Shames17}.

That said however the PL brightness is still substantial, and importantly, combined with natively high-contrast hyperpolarized $\Cs$ MR imaging that HTA-fabricated particles offer, multicolor optical imaging can allow new multimodal imaging avenues. Indeed, we envision the diamond particles being imaged under MRI as well as more than one optical wavelength simultaneously~\cite{Lv19}. This is aided by the fact that hyperpolarization can be carried out under any wavelength $\xl\lesssim$575nm. Not only can imaging therefore proceed at the two widely disparate frequencies (RF and optical) {simultaneously}, but as we showed in recent work~\cite{Lv19}, exploiting the reciprocal space nature of optical and and MR imaging can lead to substantial imaging acceleration in practical settings. In addition, both optical and MR modalities can be individually signal modulated on-demand, engendering lock-in techniques that can strongly suppress image backgrounds~\cite{Lv19}. It is in this context that the current work bears importance, since it provides a simple ``\I{materials-only}'' means to increase diamond $\Cs$ MR image signal-to-noise (SNR), while also concurrently allowing fluorescence in multiple optical channels, all of which portend approaches for high-fidelity particle tracking in-vivo~\cite{Whiting15}.

\T{\I{Conclusions:}} -- 
We have demonstrated the potential of high temperature annealing towards boosting NV-center driven $\Cs$ hyperpolarization in particulate diamonds at room temperature. At high \NV concentrations, the use of HTA at 1720\dC allows a boost in hyperpolarization levels by over an order of magnitude. We leveraged combined EPR and NMR techniques to systematically unravel origins of these surprising gains, studying DNP under several material dimensions including \NV center concentration, annealing conditions and particle size. It emerges that HTA serves to relieve radiation induced lattice disorder in the diamond lattice, reducing the triplet paramagnetic content, and increasing both the \NV and $\Cs$ $T_1$ lifetimes by factors greater than 3-5 fold each. Overall this reveals the central role played by lattice quality in determining the final saturation $\Cs$ hyperpolarization levels and suggests methods for the guided discovery of high quality nanodiamond particles for enhanced spectroscopy and MR imaging. We  anticipate that the lattice benefits from HTA will also translate to improved samples for quantum sensing and magnetometry~\cite{Lesage12,Wolf15}, especially in the limit of high NV center densities. Finally, high temperature annealing opens interesting opportunities for defect center manipulation, with potentially wider applications in quantum computing platforms in solids~\cite{Weber10,Sipahigil16}.

\T{\I{Acknowledgments}} -- It is a pleasure to gratefully acknowledge discussions with B. Gilbert and D. Suter. This material is based in part upon work supported by the National Science Foundation Grant No. 1903803. Adamas acknowledges partial support from the National Cancer Institute of the National Institutes of Health under Award No. R43CA232901 and from the NHLBI, Department of Health and Human Services, under
Contract No. HHSN268201500010C.

\T{\I{Methods:}} -- Hyperpolarization at room temperature is carried out by nine fiber-coupled 520 nm laser diodes (Lasertack) at a 38mT polarizing field generated by a pair of Helmholtz coils. A self shorted split-solenoid coaxial line connected to a 16W MW amplifier provides the microwave excitation that initially consists of frequency chirps generated from three cascaded voltage controlled oscillator (VCO) sources driven by a home-built quad ramp voltage generator~\cite{Ajoy18pol}. The diamond particles are immersed in water to leverage scattering-induced uniform illumination, as well as for heat sinking. DNP signals are measured in a  7T Oxford magnet subsequent to sample shuttling to high field. This is achieved by a home-built field cycling stage consisting of a precise belt-driven actuator stage (Parker) with 50$\mu$m precision~\cite{Ajoyinstrument18}. Shuttling times are measured to be $\app$640ms, small compared to $\Cs$ nuclear $T_1$. 

Annealing of irradiated diamond powder in standard regime (800-900\dC) was done using a Blue Lindberg Model 848 furnace. Higher temperature annealing was performed using an all-graphite furnace, model HTT-G10, MEO Engineering Company, adapted to treatment of diamond powder with a special graphite sample container (\zfr{RTA}A). Heating could be ramped up to the target temperature within minutes. 
For annealing of diamond powder at 1700-1800\dC the heating up time was about 3min, then temperature was maintained for specified time at target temperature ($\pm$10\dC) and followed by cooling down within 3min  to about 500\dC and about 5min for cooling down to room temperature. Temperature was measured using a W-Re thermocouple calibrated against melting points of pure Al, Au, Cu, Si, Ni, Pt, Pd, Rh and Mo.  In majority of experiments HTA annealing was performed in hydrogen atmosphere. For a comparison in the experiment of annealing at 1750\dC both hydrogen and vacuum had been used.

Photoluminescence measurements in \zfr{NVconc}B and \zfr{multicolor} at room temperature were performed after each step of annealing. Analysis was performed using a using an Olympus IX71 inverted fluorescent microscope and a modular USB spectrometer (HR2000, Ocean Optics). Excitation from a mercury lamp was filtered with a D350/50x band pass filter (Chroma) for UV excitation and BLP01-405R long pass emission filter (Semrock); FF01-470/28 (Semrock) bandpass filter for blue excitation measurements, with emission collected using a BLP01-488R (Semrock) long pass filter; and BLP01-532R (Semrock) bandpass filter for green excitation measurements, with emission collected using a BLP02-561R (Semrock) long pass filter. The microscope is also fitted with an 5.0 MP CCD color camera for simple color imaging applications (AmScope, MT5000-CCD-CK). Room temperature (T=295K) EPR measurements in \zfr{EPR} were carried out at using a Bruker EMX-220 spectrometer equipped with an Agilent 53150A frequency counter. 

\bibliographystyle{apsrev4-1}

\pagebreak
\clearpage
\onecolumngrid
\begin{center}
\textbf{\large{\textit{Supplementary Information} \\\smallskip
\bluetitle{High temperature annealing enhanced diamond $\Cs$ hyperpolarization at room temperature}}}\\
\hfill \break
\smallskip
M. Gierth,$^{1}$  V. Krespach,$^{1}$ A. I. Shames,$^{2}$  P. Raghavan,$^{3}$ E. Druga,$^{1}$ N. Nunn,$^{4}$ M. Torelli,$^{4}$ R. Nirodi,$^{1}$ S. Le,$^{1}$ R. Zhao,$^{1}$\\ A. Aguilar,$^{1}$ X. Lv,$^{1}$ M. Shen,$^{1}$ C. A. Meriles,$^{5}$ J. A. Reimer,$^{3}$  A. Zaitsev,$^{6,7}$ A. Pines$^{1}$, O. Shenderova$^{4}$ and A. Ajoy$^{1,8,\BRd{\ast}}$\\
\smallskip
\emph{${}^{1}$ {\small Department of Chemistry, and Materials Science Division Lawrence Berkeley National Laboratory University of California, Berkeley, California 94720, USA.}}
\emph{${}^{2}$ {\small Department of Physics, Ben-Gurion University of the Negev, Be’er-Sheva 8410501, Israel.}}
\emph{${}^{3}$ {\small Department of Chemical and Biomolecular Engineering, and Materials Science Division Lawrence Berkeley National Laboratory University of California, Berkeley, California 94720, USA.}}
\emph{${}^{4}$ {\small Adámas Nanotechnologies, Inc., 8100 Brownleigh Dr, Suite 120, Raleigh, NC, 27617 USA.}}
\emph{${}^{5}$ {\small Department of Physics and CUNY-Graduate Center, CUNY-City College of New York, New York, NY 10031, USA.}}
\emph{${}^{6}$ {\small College of Staten Island, CUNY, 2800 Victory Blvd., Staten Island, New York 10312.}}
\emph{${}^{7}$ {\small Gemological Institute of America, 50 W 47th, New York, NY 10036 USA.}}
\emph{${}^{8}$ {\small Department of Chemistry, Carnegie Mellon University, 4400 Fifth Avenue, Pittsburgh, PA 15213, USA.}}
\end{center}



\twocolumngrid

\beginsupplement

\begin{figure}[t]
  \centering
  {\includegraphics[width=0.45\textwidth]{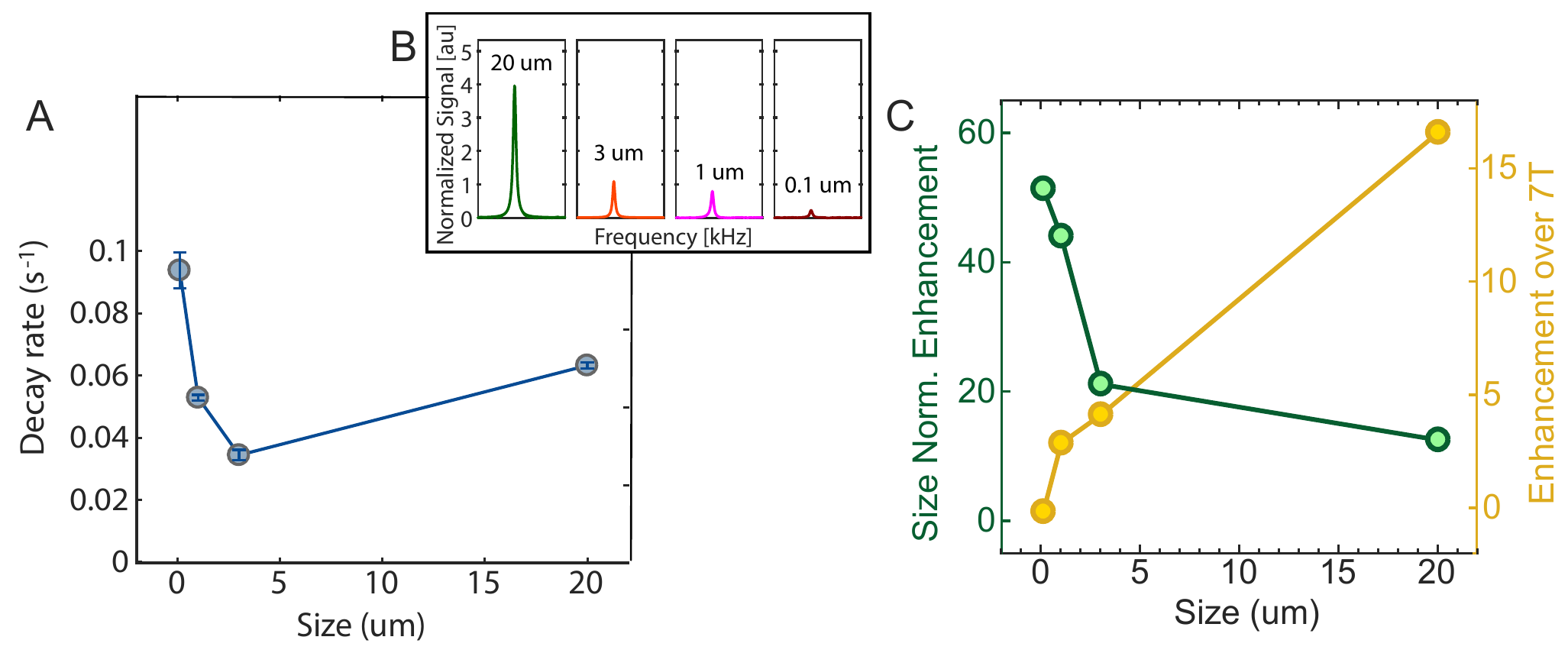}}
\caption{\T{DNP dependence on particle size} studied for samples fractionally milled starting from the same parent material. (A) Polarization decay rates at 38mT reveal a weak dependence, and indicate that the hyperpolarization decay is approximately independent of particle size down to 100nm. The 3$\mu$m sample has a longer $\Cs$ lifetime since a lower electronic irradiation energy was employed in its production. (B) Representative mass normalized $\Cs$ hyperpolarized spectra obtained from the particles of varying size.  (C) Mass normalized polarization enhancements over 7T (yellow line) show a steep decrease in the DNP efficiency with particle size.  However, when the data is additionally normalized with respect to surface area-to-volume ratio (green line), the smaller particles down to 100nm show the best overall hyperpolarization levels per number of surface $\Cs$ nuclei. }
\zfl{size}	
\end{figure}

\section{Size dependence of hyperpolarization enhancements}

We consider in \zfr{size} the dependence on particle size, employing particles starting from the same parent material that have been fractionally milled to sizes as low as 100nm nanodiamonds NDs. Hyperpolarization in small particles is important both for applications employing them as agents for optical DNP of external liquids, as well as for MRI imaging of the particles themselves. This is on account of their increasing surface area to volume ratio that scales linearly with decreasing particle size, and the fact that small NDs can be safely injected into (and ejected from) \I{in-vivo} target disease locations. 

We find the DNP enhancements decreasing with particle size (see \zfr{size}B), which we speculate to be arising from the increased role of crushing related surface effects at smaller particle sizes. There is also NV$^{-}$/ NV$^0$ stronger charge dynamics at play for smaller NDs under under optical pumping. When normalized by the effective particle surface area to volume ratio however, the 100nm ND sample provides the best overall hyperpolarization efficiency, pointing that they are still most suitable amongst other particles as an agent for external polarization. Moreover, we observe no direct correlation in the spin relaxation lifetimes with size at least down to 100nm. The enhanced $T_{1n}$ lifetime of the 3$\mu$m particles in \zfr{size}A were because although the same electron fluence was used in all samples, this sample had a lower electron irradiating energy (1MeV in contrast to 3MeV in all other samples). This substantiates the notion that longer irradiation times at lower doses result in lower lattice damage and consequently high $\Cs$ relaxation times.

\begin{figure}[t]
  \centering
  {\includegraphics[width=0.45\textwidth]{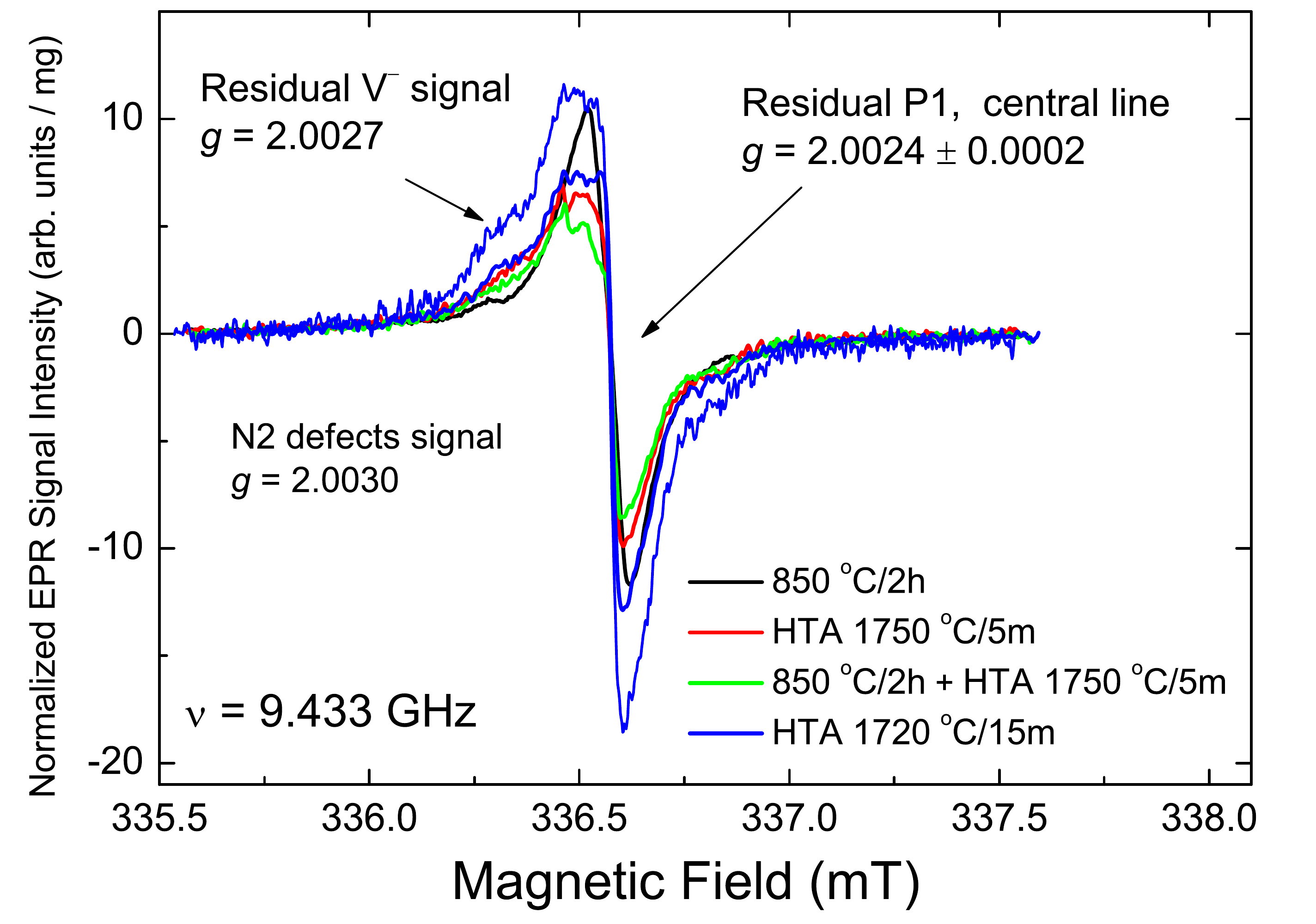}}
\caption{\T{EPR spectra of primary defects with and without HTA.} Room temperature X-band EPR spectra of $D_2$= 5$\zt$10$^{19}$e/cm$^2$ samples with and without different classes of HTA (see also \zfr{EPR}C). Here we concentrate on the $g$ = 2.00 region (primary defects), central components, and varying annealing conditions are indicated in the legend. EPR parameters employed in these measurements were: P$_{\R{MW}}$ = 2$\mu$W, A$_{\R{mod}}$ = 0.005 mT, RG = 2$\zt$10$^5$, n$_{ac}$ = 100, and frequency $\nu$ = 9.433 GHz. Differences in signal-to-noise ratios appear after normalization to different masses of the samples.}
\zfl{primary_EPR}	
\end{figure}

\begin{figure}[t]
  \centering
  {\includegraphics[width=0.45\textwidth]{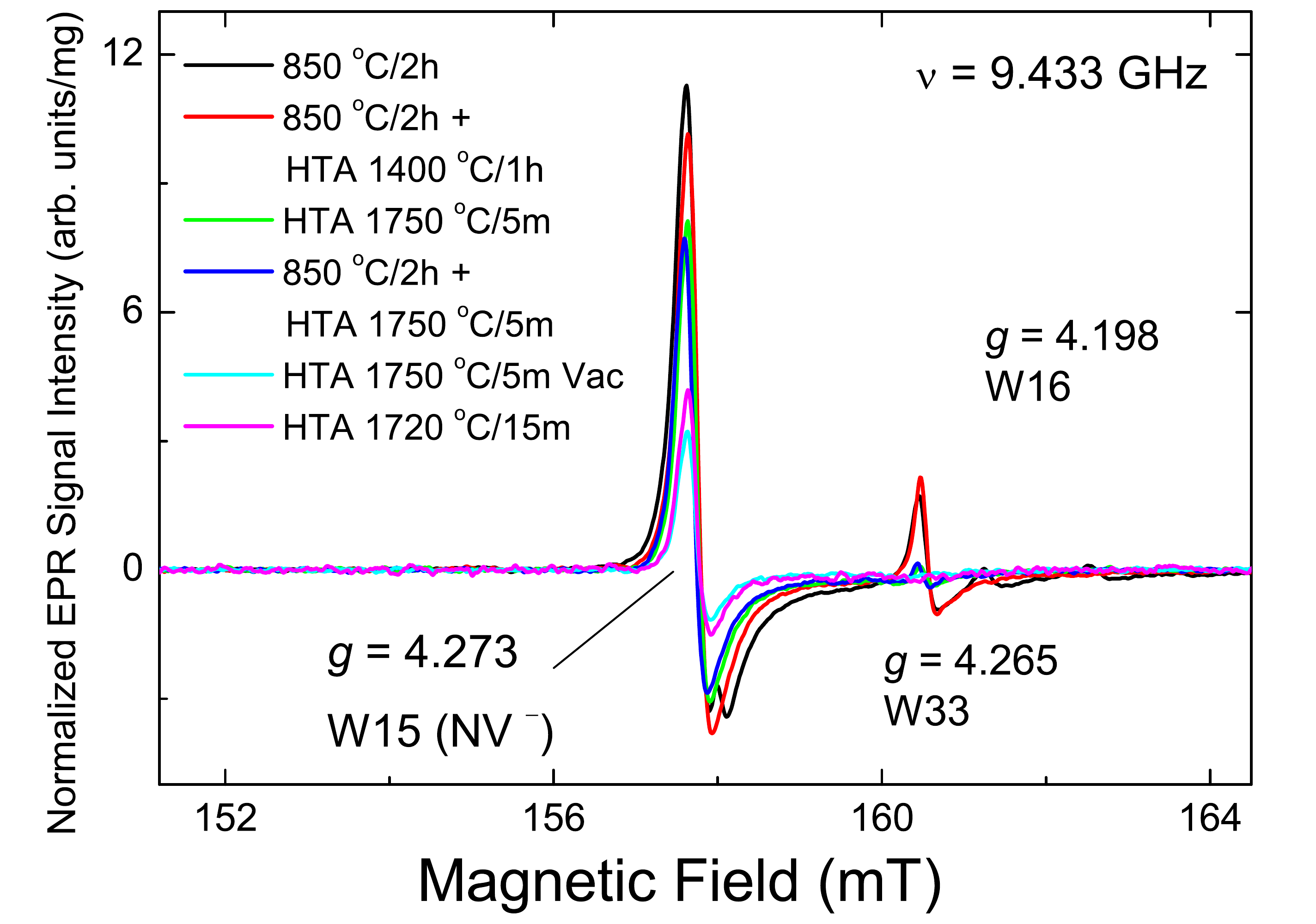}}
\caption{\T{EPR spectra of triplet defects with and without HTA.}  Room temperature X-band EPR spectra of $D_2$= 5$\zt$10$^{19}$e/cm$^2$ samples with and without different classes of HTA (see also \zfr{EPR}C). Here we concentrate on the spectral region close to the NV centers and triplet defects W16-W33 (see also \zfr{EPR}B). EPR parameters employed in these measurements were: P$_{\R{MW}}$ = 100$\mu$W, A$_{\R{mod}}$ = 0.075 mT, RG = 2$\zt$10$^5$, n$_{ac}$ = 256, and frequency $\nu$ = 9.433 GHz.}
\zfl{triplet_EPR}	
\end{figure}

\begin{figure}[t]
  \centering
	\begin{subfigure}
  {\includegraphics[width=0.44\textwidth]{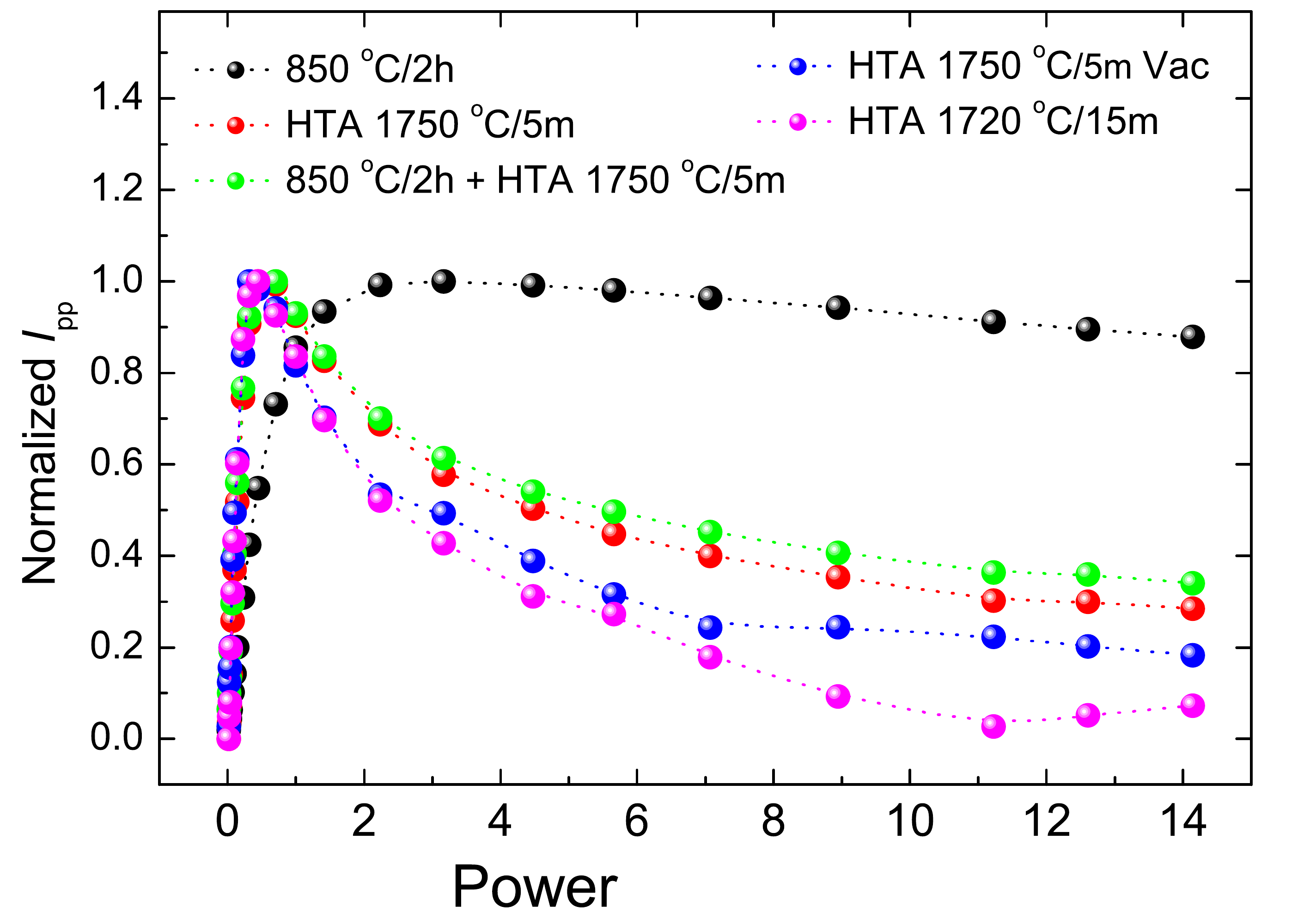}}
	\end{subfigure}
	
	\begin{subfigure}
  {\includegraphics[width=0.44\textwidth]{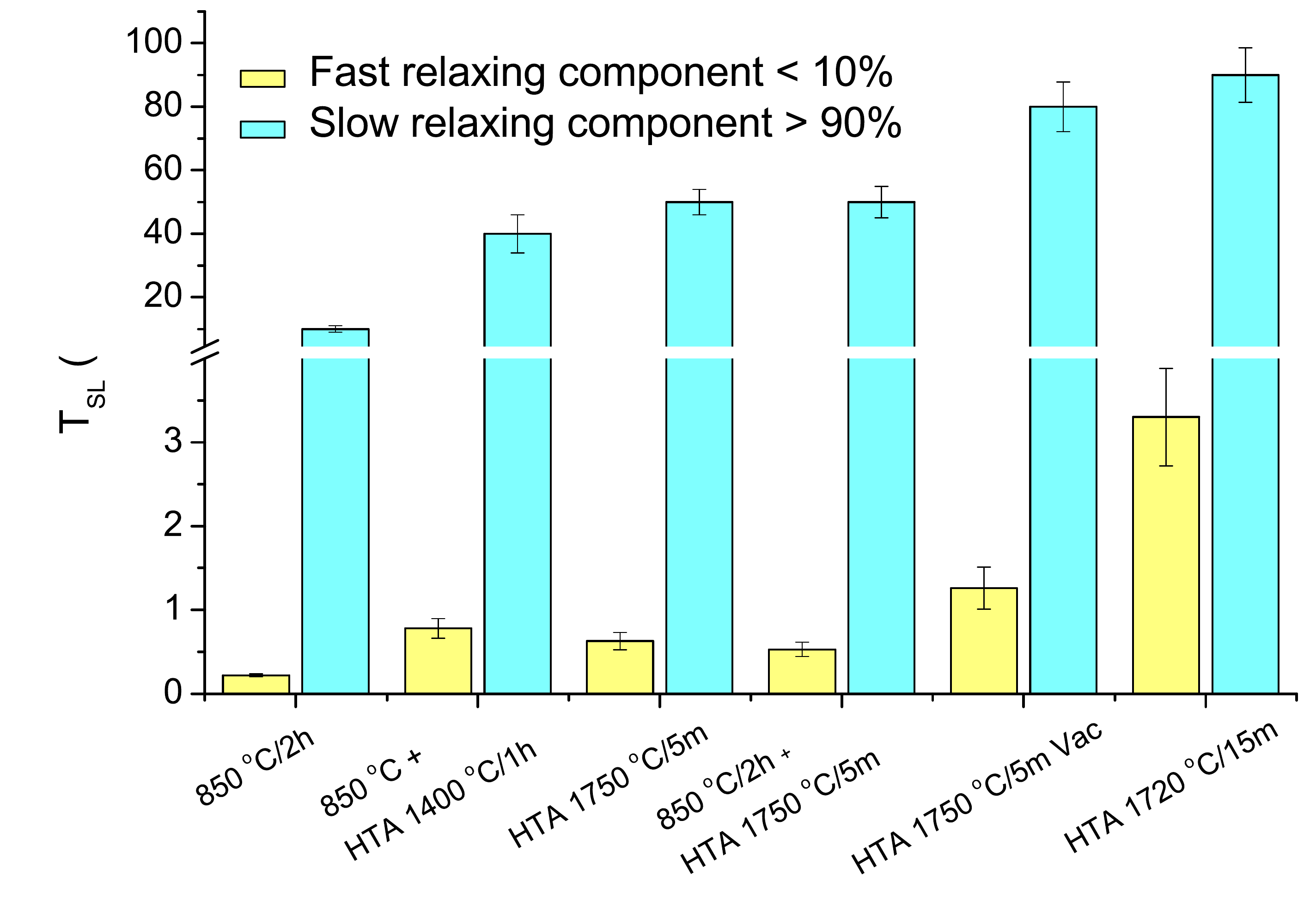}}
	\end{subfigure}
\caption{\T{EPR data accompanying \zfr{EPR}E}. \I{Upper panel:} Relaxation curves for samples annealed at various conditions measured on the “forbidden” g = 4.273 line of the \NV spectrum. \I{Lower panel:} electron spin-lattice relaxation times estimated for EPR lines attributed to the  “forbidden” transitions between Zeeman sublevels of the triplet \NV- centers. }
\zfl{forbidden_EPR}	
\end{figure}

\section{Additional EPR data}
We provide here the raw data of the EPR measurements that form the basis of \zfr{EPR} of the main paper. Consider first \zfr{primary_EPR}, which displays the EPR spectra of primary defects with and without high-temperature annealing, and from which the concentrations in \zfr{EPR}D are extracted. It is evident that HTA has an impact on the primary defect concentration, increasing the P1 concentration, while slightly decreasing the concentrations of V$^{\text{-}}$ and N3 defects.

Similarly \zfr{triplet_EPR} contains the raw data of the EPR spectra in the spectral range close to the \NV center, and showing the triplet defect centers. It is here that the effective HTA is most pronounced: there is a steep decrease in the triplet defect center concentration (W16 ,W33) while the \NV concentration is only weakly affected. Finally, \zfr{forbidden_EPR} serves as a complement to the EPR microwave saturation curves displayed in \zfr{EPR}E-F of the main paper, but performed here for the forbidden transition between Zeeman sublevels of the \NV centers. The increase in the effective \NV spin-lattice relaxation time is even evident, and we separate the slow and fast relaxing components assuming a biexponential fit in the top panel of \zfr{forbidden_EPR}.

\section{Experimental Apparatus}
We now provide more details on the hyperpolarization experimental setup. We refer the reader to a detailed exposition in Refs.~\cite{Ajoy17, Ajoy18pol}, but sketch the apparatus here for completeness. Generating bulk polarization in the diamond powders relies continuous laser pumping at low-fields, and subsequent sample transfer to high field (7T) magnet for NMR inductive readout. Transport of the samples is controlled by a rapid motion actuator (Parker HMRB08) with precision $\pm50 \mu$m from point-to-point. Actuator software allows for a programmable S-curve motion profile, with the device moving up along the magnetic field axis with 2m/s velocity and 30 m/s$^2$ acceleration for a length of approximately 1.6m. The transport system was tested to confirm its consistency over more than 1400 trials, finding a mean transport time of 648$\pm$2ms between the polarization of the sample under the magnet and NMR detection. $\Cs$ polarization is detected at 75.03MHz by printed copper NMR saddle coils laid around a quartz tube. The diamond powder samples were placed into the bottom of NMR tubes (Wilman 8mm OD, 1mm thickness), and then filled to about three quarters the length with distilled water. 

To maximize the surface area exposed to laser irradiation during optical pumping, an octagonal 8-laser system has been designed (see \zfr{laser}E) composed of fiber optic cables (Thorlabs M35L01) that emit the 532nm radiation from diode lasers assembled below the polarization region. These cables are arranged symmetrically onto an octagonal “\I{puck}” structure with the DNP microwave coil at the center, allowing synchronous application of microwaves with this improved laser excitation design. A ninth fiber optic cable is additionally fed below the sample. The puck design allows for stacking of identical multi-laser devices that will be useful for proportional gains in samples of diamond powder of greater total mass, with applications in using larger total surface area to transfer polarization to liquids. 

In an effort to characterize the uniformity of the obtained hyperpolarization in the sample, we perform in \zfr{betagamma} experiments similar to \zfr{laser}F of the main paper employing identical 200$\mu$m diamond particles (Element6), but with differing masses. We consider the limit of low and high masses (8mg, 40 mg) respectively. The nearest neighbor laser hyperpolarization charts in \zfr{betagamma}A, and the accompanying scaling of the hyperpolarization with the number of lasers (\zfr{betagamma}C-D) show a faster saturation for the lower mass. We note that compared to \zfr{laser}F, the larger particles employed here have a lower optical penetration, but even for these large particles masses of $\lesssim$20 mg can be almost completely optically polarized. Importantly in \zfr{betagamma}E, we unravel the effect of each laser on the hyperpolarization process, by considering the \I{correlation} of the obtained hyperpolarization enhancements in the case of employing two lasers that are not necessarily nearest neighbor only. In particular, we employ laser numbered $k_j$ and laser $k_{j+n}$, separated along the octagonal ring simultaneously. We compare the effect on the hyperpolarization signal $\mC_{k_j,k_{j+n}}$ to the sum of the signals, $(\mC_{k_j} + \mC_{k_{j+n}})$, obtained applying the two lasers separately. The normalized result, $\expec{C_{k,k+n}} = \fr{1}{8}\sum_j \mC_{k_j,k_{j+n}}/(\mC_{k_j} + \mC_{k_{j+n}})$, averaged over all $k_j$ on the ring are displayed in \zfr{betagamma}E. The results indicate that, to a very good approximation, the buildup of polarization is additive between disjoint laser pairs, and indicative of macroscopically uniform polarization buildup within the sample.

\begin{figure}[t]
  \centering
  {\includegraphics[width=0.45\textwidth]{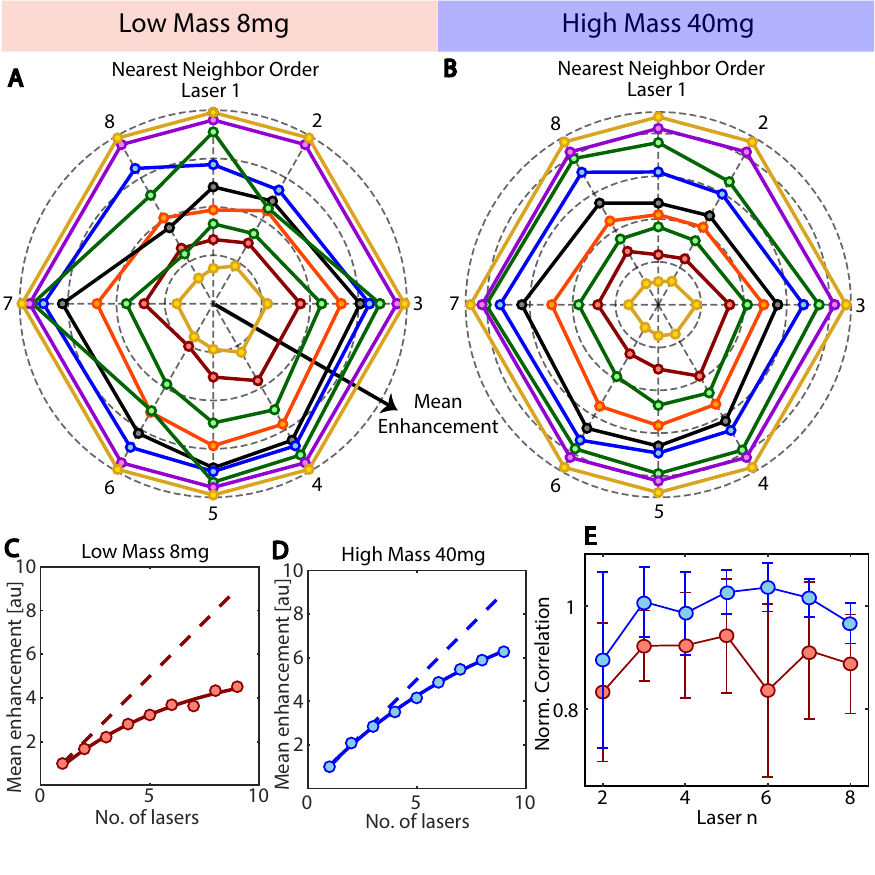}}
\caption{\T{Spatial hyperpolarization buildup charts} for (A-B) nearest neighbor laser combinations (see \zfr{laser}F) for two masses (8mg, 40mg) of the same sample (200$\mu$m microparticles). (C-D)  Corresponding scaling of the mean enhancement with increasing nearest neighbor laser power. (E) Comparison of the hyperpolarization laser correlation $\expec{C_{k,k+n}}$ for the two samples considered. Panel illustrates that the polarization buildup, to a large extent, is additive between disjoint laser pairs.
}
\zfl{betagamma}	
\end{figure}

\section{Data processing}

In this section we discuss the experiment protocols used to generate buildup and decay curves (see Fig 2 of main paper) for the various diamond samples. The buildup curves are obtained through normal operation of the hyperpolarization apparatus and field cycler as described in the previous section, with sole variation in the duration of hyperpolarization time (laser pumping time). Decay curves are similarly generated by now polarizing for a fixed time and adding a variable wait time before NMR detection. Sweeping of the \NV $m_s = +1$ manifold at 200Hz and 36mT is maintained synchronously with 520nm laser irradiation for all polarization detection experiments to maximize $\Cs$ polarization. 

Next we discuss the evaluation and signal enhancement calculation of our NMR data, which is done through a phasing and fitted enhancement calculation. The data is first phased through a zero-order correction that attempts to minimize the imaginary value parts of the peak, and maximize the real values. This is done through multiplication by a phase value, and this value also determines the sign of the peak (positive or negative). Because the zero-order phasing does not always fix the entire spectrum, the data is then baseline-corrected with a 12th order polynomial. A Lorentzian is fit through the entire data, and the peak limits are defined such that they enclose 90\% of the total Lorentzian curve area; data on the sides of the peak limits are considered the baseline, which is then flattened by subtracting the polynomial from the phased baseline values. Finally, the NMR spectrum is compared to a thermal spectrum by scaling the noise of both to 1, and evaluating the enhancement as follows for differing number of averages of the DNP and thermal signals $N_{\R{DNP}}$ and $N_{\R{Thermal}}$ respectively

$$ \epsilon = \frac{\R{SNR}_{\R{DNP}}}{\R{SNR}_{\R{Thermal}}}\sqrt{\frac{N_{\R{DNP}}}{N_{\R{Thermal}}}} $$.

\subsection{$T_1$ Data Processing}

Let us now consider details of the experiments mapping the $\Cs$ relaxation times in diamond as a function of magnetic field. We refer the reader to a more detailed exposition in Ref.~\cite{Ajoy19relax}. $T_1$ relaxation of the nuclear spins arise from spin-flipping processes that randomly flips their nuclear spin, for which the Hamiltonian is of the form,
\beq
\mH_I = \xo_LI_{z} + \expec{A_{zx}}s_z(t)I_{x}\:.
\zl{fluc}
\eeq
where $s(t)$ is a stochastic variable. This Hamiltonian can arise in an effective rotating frame, for instance due to the coupling of $\Cs$ spins to a reservoir of dipolar coupled paramagnetic spins in its vicinity. 
In a highly simplified picture, pairs of paramagnetic spins spatially remote to the $\Cs$ nucleus and undergoing spin flip-flops can flip the $\Cs$ spin if it’s energy $\xo_L=\xg_nB$ determined by the magnetic field is within the dipolar coupling spectrum of the bath electrons, i.e. $\xo_L\leq \expec{d_{ee}}$. In previous work we had demonstrated this electron mediated process as being the dominant relaxation mechanism at low-moderate fields, and operational at fields under which hyperpolarization is carried out. 
Measuring the field profile $T_1(B)$ can spectrally fingerprint the processes underlying the relaxation. To carry out these experiments, we interface the hyperpolarization setup with a precision field cycling instrument. At every field point in the 100mT-3T range, a relaxation curve is obtained in order to extract the effective $T_1$ relaxation time for the spin that field. 
Perhaps most illustrative in the data is that samples with prepared by RTA conditions show a much narrower $T_1(B)$ field profile. This directly translates to the fact that at any field location, the relaxation time of the sample is much longer. Consequently, there are larger DNP enhancements,since the polarization can buildup in the sample for a longer period. 

As an outline of the steps in the processing of the relaxation data, the signal enhancement as a function of time can be described using the following equation:
\begin{equation}
	\centering
	\epsilon(t) = \epsilon_o e^{-\frac{t}{T_1}}.
\end{equation}
where $\epsilon_o$ represents the initial enhancement before any signal decay is observed. Values for $\epsilon(t)$ for the two samples in \zfr{RTA}C were obtained by first optically pumping the samples for 60 seconds and measuring their decay over a time span of 30 seconds for a range of fields between 0 and 3500 Gauss. For each set field value, the enhancement after 30 seconds of depolarization was measured 10 times and then averaged. The initial enhancement value $\epsilon_o$ was obtained by again pumping each sample for 60 seconds and taking the enhancement value after zero waiting time. This process was repeated 30 times and the resulting $\epsilon_o$ values were then averaged. Once the initial enhancement and field dependent enhancement values were obtained, the relaxation rates $R_1$ could be calculated by rearranging the previous equation:
$R_1 = \frac{1}{T_1} = \frac{ln\big(\frac{\epsilon_o}{\epsilon(t)}\big)}{t}\:$.

The sum of two Tsallian functions was then constructed to fit the relaxation data (solid lines in \zfr{RTA}C). This is described by the following equation:
\bea
    R_1(B) &=& A_1\bigg[\big(1+2^{q_1 - 1}\big)\lb\frac{B}{w_1}\rb^2\bigg]^{-\frac{1}{q_1 - 1}}\non\\
		&+& A_2\bigg[\big(1+2^{q_2 - 1}\big)\lb\frac{B}{w_2}\rb^2\bigg]^{-\frac{1}{q_2 - 1}} + c
\eea
where the fit parameters $A_1$ and $A_2$ represent the amplitudes, $w_1$ and $w_2$ the widths, $c$ represents the vertical offset, and $q_1$ and $q_2$ tell us if the function more closely represents a Lorentzian or a Gaussian. A $q$ value of 1 indicates a perfect Lorentzian fit and a q value of 2 indicates a perfect Gaussian. These five fit parameters were then allowed to vary until they were optimized.

\section{Injection Rate Calculations}
We define the injection rate as the rate at which the hyperpolarization is injected from the \NV centers spins to the $^{13}$C reservoir. The efficiency of this hyperpolarization process depends on several factors such as \NV center coherence time, laser penetration, and the effective NV-$^{13}$C hyperfine couplings. The effective polarization injection is an average effect over the entire sample. While the injection rate is reflect in the polarization buildup curves that we measure, these curves also include the inherent polarization decay in the samples. To a good approximation, the injection rates can be quantified as the effective difference between the DNP buildup and decay curves at any field. 

We can estimate the effective polarization buildup and decay rates by assuming a monoexponential growth and decay and constructing the following differential equation,
\begin{equation}
    \centering
    \frac{d\epsilon(t)}{dt} = -\beta_1\epsilon + \beta_2[P_o - \epsilon]
\end{equation}
where $\epsilon$ is the enhancement as a function of time, $P_o$ is the initial polarization, $\beta_1$ is the decay rate, and $\beta_2$ is the injection rate. Solving this differential equation and applying the initial condition $\epsilon(t=0) = 0$ since we note that the average polarization can not build up or decay instantaneously, we find:
\begin{equation}
    \centering
    \epsilon(t) = \frac{P_o\beta_2}{\beta_1 + \beta_2}\big[1-e^{-(\beta_1 + \beta_2)t}\big]
\end{equation}
Since the rates in practice are actually bi-exponentials, we evaluate the effective rate constants by finding their 1/e crossing times. This allows us to calculate the effective polarization injection rates as well as the error bars arising from the difference between the two exponentials.

\end{document}